\documentclass{aa}
\usepackage{graphicx}
\usepackage{natbib}
\usepackage{psfig}
\usepackage{amsmath}
\usepackage{aas_macros}
\usepackage{natbib}
\bibpunct{(}{)}{;}{a}{}{,}

\def\v{~variability~}
\usepackage[center]{subfigure}

\begin{document}
\title{X-ray spectral and timing characteristics of the stars in the young open cluster IC
2391} 
\author{A. Marino\inst{1,2}
\and
G. Micela\inst{2}
\and
G. Peres\inst{1}
\and
I. Pillitteri\inst{1}
\and
S. Sciortino\inst{2}
}

\offprints{A. Marino  e-mail: marino@astropa.unipa.it}

\institute{Dipartimento di Scienze Fisiche \& Astronomiche, Sezione di Astronomia, Universit\`a di Palermo, Piazza del Parlamento 1, 90134 Palermo - ITALY;
\and
INAF - Osservatorio Astronomico di Palermo
Piazza del Parlamento 1, 90134 Palermo - ITALY;
}

\date{Received 24 June 2004 / accepted 19 September 2004 }

\abstract{We present X-ray spectral and timing analysis of
members of the young open cluster IC 2391 observed with
the XMM-Newton observatory.
We detected  99 X-ray sources by analysing the summed  data obtained 
from MOS1, MOS2 and pn detectors of the EPIC camera;
24 of them are members, or probable members, of the cluster. 
Stars of all spectral types have been detected, from the early-types 
to the late-M dwarfs. 

Despite the capability of the instrument to recognize up to 3 thermal 
components, the X-ray spectra of the G, K and M members of the cluster 
are well described with two thermal components 
(at kT$_1 \sim$ 0.3-0.5 keV and kT$_2 \sim$ 1.0-1.2 keV respectively) 
while the X-ray spectra of F members require only a softer 1-T model.  

The Kolmogorov-Smirnov test applied to the X-ray photon time series shows 
that approximately 46\% of the members of IC 2391 are variable with a 
confidence level $>$99\%.
The comparison of our data with those obtained with ROSAT/PSPC, nine
years earlier, and ROSAT/HRI, seven years earlier, shows that
there is no evidence of significant  \v on these time scales,
suggesting  that long-term variations due to activity cycles similar to that on the 
Sun are not common, if present at all, among these young stars.
 }

\titlerunning{X-ray characteristics of IC 2391 stars}
\authorrunning{Marino et al.}

\maketitle
\section{Introduction}
Most of our knowledge on the long-term evolution of coronal emission derives 
from observations of open clusters, which supply large, chemically 
homogeneous, and well dated samples.  First X-ray studies of open clusters 
with the {\em Einstein} Observatory \citep{Stern81, Cai85, 
Mice85, Mice88, Mice90} showed that the average level of X-ray 
luminosity decays with age.
ROSAT observations have extended the {\em Einstein} results both by
enlarging the number of  known X-ray emitters in a given cluster and covering
clusters over a much  wider age range \citep[e.g.,][]{Ster95, Gag95,
Mice96, Mice99, Stau94, Ran95, Ran96, Pro96, Jeff97, Ra98, Mice00, Fra00,Scio00, Bar02}. 
The comparison of different open clusters shows that the X-ray emission
decreases from younger clusters (like IC 2391, IC 2602 at $\sim$ 30 Myr) 
to intermediate age clusters (like $\alpha$ Per, Pleiades and NGC2516 at 
$\sim$ 50 - 100 Myr), to older clusters (like Hyades, Coma, Praesepe at 
$\sim$ 500-700 Myr).  There is ample consensus that X-ray luminosity
evolves during the stellar lifetime mainly as a result of the spin down due to 
magnetic braking.
The coronal emission is influenced by several stellar properties 
but the rotation, in particular, plays a fundamental role; since 
rotation evolves with age so does coronal activity. 

Data on open  cluster have shown that the most  active stars, with rotation velocities
above $\sim$ 15-20 km/s, reach a maximum X-ray luminosity such that 
L$_{{\rm x}}$/L$_{{\rm bol}}$ saturates at $\sim$ 10$^{-3}$,
where L$_{\rm bol}$ is the star's bolometric luminosity \citep[e.g.][]{Vi84, Piz03}.
Saturation lacks a clear  interpretation: it could be the effect
of dynamo saturation, or it could correspond to the total filling  of the star's surface
by active regions, as originally suggested by \citet{Vi84} or it could be the result of 
other, yet unknown, phenomena.
Furthermore ROSAT observations  have shown that stars rotating even 
faster ({\em  v~sin~i~}$>$ 100  km/s) exhibit a level of X-ray luminosity  
3-5 times below the saturation level \citep{Ra98}. 
The origin of this phenomenon, named supersaturation by \citet{Pro96},
is still unexplained and several hypotheses have been formulated. Supersaturation could
be the result of an overall decrease of dynamo efficiency at very high rotation rates 
or the consequence of the redistribution of the radiative output due to the lower 
stellar "effective" gravity because of the enhanced rotation (Randich 1998). 
Recently, \citet{Mar03} has found evidence of rotational modulation in a 
supersaturated star of IC 2391 with a very short photometric period of 0.22 days. 
The observed modulation implies that the star is not completely covered by active regions.
As an explanation, \citet{Jar04} has suggested that the X-ray emitting corona 
of rapid rotators suffers a centrifugal stripping of coronal
loops, producing open field regions that become dark in X-rays.\\
The IC 2391 open cluster is an ideal laboratory for testing activity-age
relations in fast rotators, because it is young enough
($\sim$ 30 Myr) that for solar-type members to have just arrived on the zero-age
main-sequence (ZAMS). 
At a  relatively nearby distance (162 pc), this  southern cluster 
contains about 172 known members or probable members, ranging in spectral
type from B to M, enabling the simultaneous study of the different processes driving X-ray
emission from stars of different internal structure.
The first X-ray observation of the cluster was performed with the ROSAT/PSPC 
\citep{Pa93, Pa96}, followed by a 56 ksec ROSAT/HRI observation \citep{Si98}.
These observations have shown that among the late-type members of the 
cluster  the spread of X-ray luminosities is as large as a factor $\sim $20. 
A similar spread is observed in the rotational periods, implying that the stars 
arrive on the ZAMS with a wide range both in angular momentum and activity level values.
However, the limited sensitivity of the ROSAT observations  has allowed the
detection of only a small fraction of dM cluster members. 
On the other hand some dM stars in IC 2391 have been observed and catalogued 
only recently.

In this paper we report the results of the analysis of the XMM-Newton observation
of IC2391, which covers a field of 30${\arcmin} \times 30{\arcmin}$ in the central 
region of the cluster.
The XMM-Newton/EPIC higher sensitivity and  spectral resolution, 
combined with its larger spectral band and  more continuous time
coverage with  respect  to that of ROSAT,  allow us to improve our knowledge of
spectral and timing characteristics of the cluster coronae.
Unfortunately the cluster size is $\sim$ 4 square degrees, therefore our
observation covers only a small fraction of the known members.

The structure of the paper is the following:  Sect. 2 
describes the  XMM-Newton/EPIC observation  and the data  analysis; 
Sect. 3 and 4 report  the spectral and time analysis, respectively; 
Sect. 5 summarizes the main results.

\section{X-ray observation and data analysis}
\subsection{XMM/EPIC Observation and data reduction}

We present the analysis of the Guaranteed Time XMM-Newton/EPIC observation pointed 
on the young open cluster IC 2391. 
The observation (Obs. Nr. 0112420101), centered on  $\alpha$ = 8$^{h}$ 42$^{m}$ 00$^{s}$,
$\delta$ =  -53$^{o}$ 00${\arcmin}$ 36${\arcsec}$, 
was performed on Nov 20, 2001 during  orbit 357 of XMM-Newton \citep{Ja01}. 
XMM-Newton carries three X-ray telescopes  observing simultaneously the same 
region of the sky; the collected photons form  images on three CCD-based detectors: 
the pn \citep{Str01} and the twin MOS1 and MOS2 \citep{Tur01}
which constitute the EPIC camera.  
All three EPIC detectors were active at the time  of the observation with the medium  
filter in the EPIC/pn and thick filters in the EPIC/MOS detectors.
An exposure time of 50 ksec was scheduled, but due to an intense background 
level (see below), only $\sim$ 31.4 ksec were effectively obtained in the EPIC/pn 
and $\sim$ 43.3 ksec in each of the two MOS. 

\begin{figure}
\centerline{\psfig{figure=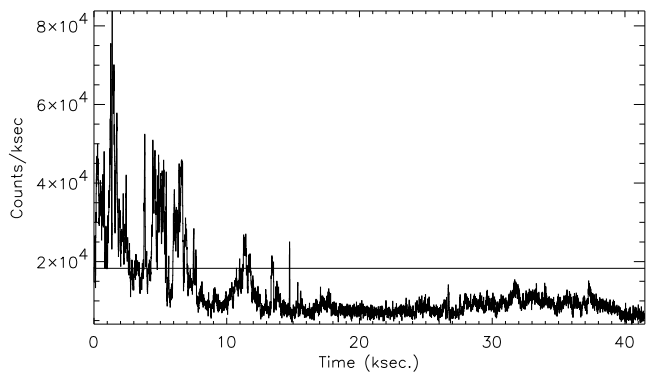,width=9.0cm}}
\caption{The overall count rate of the pn camera vs. time during the observation. 
The horizontal line is the computed threshold value for background filtering: 
time intervals during which the pn count rate was above the line have been discarded 
from the final data set.} 
\label{SN}
\end{figure}

The data have been processed using the XMM-Newton Science Analysis System (SAS)
version 5.4.1\footnote{released on January 2003 and available at the 
http://xmm.vilspa.esa.es.}.
We used the {\em epchain} and {\em emchain} tasks  to process the EPIC Observation Data
File data, obtaining three photon lists with time, position and energy of the events
recorded in the pn and MOS detectors. In order to minimize the
background due to non-X-ray events we have retained  only events  
in the 0.3-7.8 keV band  and  single, double, triple, and quadruple pixel events. 
We have limited the energy band to 0.3-7.8 keV since data below 0.3 keV are mostly
unrelated to bona-fide X-rays, while above 7.8 keV only background counts are
present for the kind of sources we are interested to.
Furthermore we have filtered the data to minimize the so-called proton flare phenomenon
which produces an enhancement of noise due to protons "focused" by XMM-Newton mirrors  and
essentially indistinguishable from bona-fide X-ray events. 
Since our aim was to maximize the signal to noise ratio, we have applied a technique 
developed at  INAF - Osservatorio Astronomico of Palermo \citep{Scio02}
that maximizes the statistical significance of weak sources by
identifying and removing fractions of the exposure time strongly affected
by  high-background episodes.
Fig. \ref{SN} shows the total count rate for the pn data set versus time,
together with the threshold (horizontal line) above which the data are discarded. 
In the observation, a negligible number of  narrow temporal segments were affected by 
a high count rate for a total of 30 s and 50 s for MOS1 and MOS2, respectively, while in 
the  EPIC/pn (see Fig. \ref{SN}) we have eliminated $\sim$ 5.4 ksec at the beginning of the 
observation. 

\subsection{Source detection}
\begin{figure}[!t]
\centering
\centerline{\psfig{figure=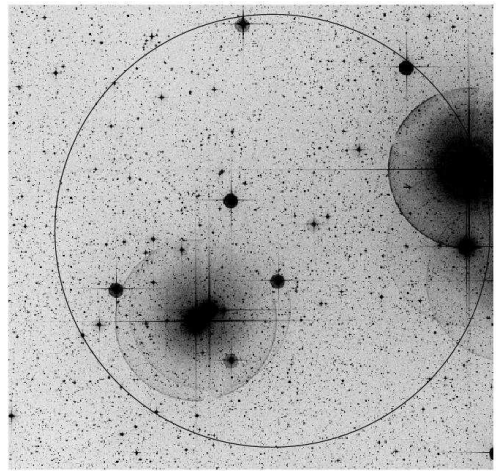,width=6.5cm}}
\centerline{\psfig{figure=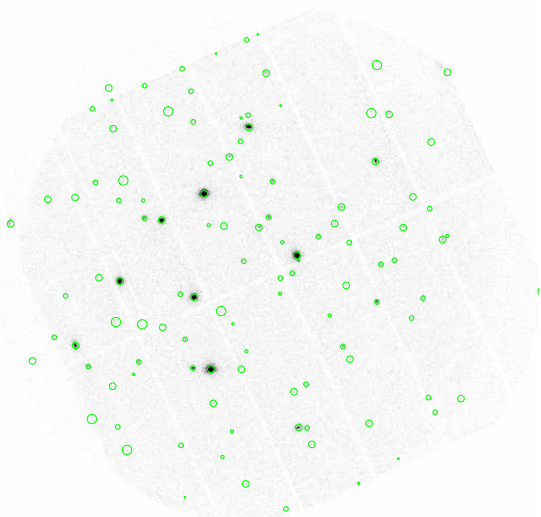,width=7.7cm}}
\caption{Top panel: the EPIC's 30 arcmin diameter FOV in the IC 2391 
Digitized Sky Survey (DSS) optical image. Bottom panel: pn, MOS1 and MOS2 
X-ray image of our  EPIC field. Circles mark the detected sources, 
(radii correspond to  the wavelet detection scales).}
\label{dss}
\end{figure}

\begin{figure}
\centering
\centerline{\psfig{figure=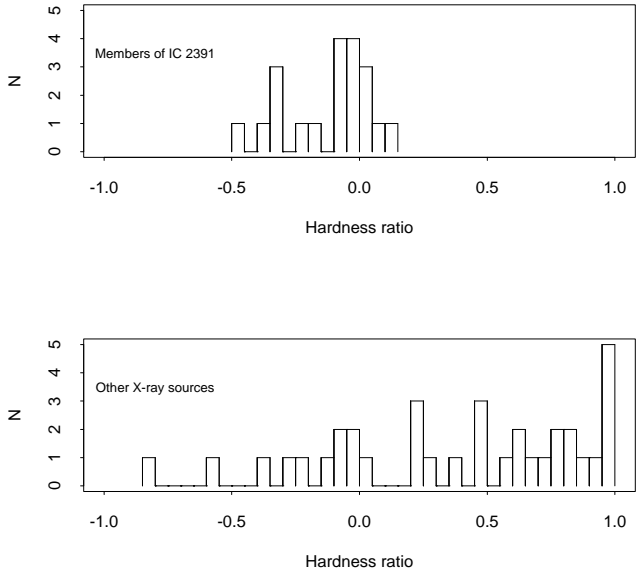,width=9.0cm}}
\caption{Histogram of hardness ratio for pn-detected members of IC 2391 (top panel) and
the other X-ray sources in the FOV (bottom panel). 
The hardness ratio is computed as (H-S)/(H+S) with H and S the counts in the 
hard (0.8-3.5) keV and soft (0.3-0.8) keV bands, respectively.}
\label{hist}
\end{figure}

The source detection has been performed on the summed data set images 
using the Wavelet Detection Algorithm developed at the  INAF - Osservatorio Astronomico di
Palermo and based on the algorithm previously developed for the ROSAT/PSPC 
\citep{Da97,Dab97} and adapted to the XMM-Newton case. 
Large sets of simulations of pure background were performed in order to derive
the appropriate detection threshold that limits the number of spurious detections. 
We adopted a threshold which statistically retains only one spurious source per field. \\
In order to perform this analysis, an exposure map was created with the SAS 
task {\em eexpmap} for each EPIC detector.
Since the pn is more sensitive than the MOS detectors, the pn time has 
been scaled  with an  appropriate factor in order to obtain a final exposure 
map in "MOS-equivalent" time.
The factor has been evaluated as the ratio between the rate measured in the pn
image and in one of the MOS images for the common sources.
The final effective time depends on the position in the field of view: 
its value  on-axis is about six times that at the edge of the field
of view; such a big difference implies a very different sensitivity for source detection 
at the center and at the edge  of the field of view (FOV). \\
We have found  99 X-ray sources (see Fig. \ref{dss} and Table \ref{Det})
after removing some spurious detections due to hot pixels not listed in the 
calibration files  we had available.
Exposure time and count rates in Table \ref{Det} are expressed in MOS1-equivalent 
units; col. 7 gives the probability level for the source to be a background 
fluctuation expressed in  terms of a gaussian-$\sigma$-equivalent.\\
We have computed  the Hardness Ratios (HR) of the X-ray sources detected in the pn
data, in order to  estimate approximately the characteristic temperature of the
emitting coronal plasma. 
The HR, evaluated as (H-S)/(H+S) (H and S are the counts in the hard (0.8-3.5 keV)
and soft (0.3-0.8 keV) bands, respectively),  is reported in col. 8 of Table \ref{Det}
with the corresponding  error computed assuming Poisson statistics in each band. 
Source photons were extracted from circular regions of 45$^{\arcsec}$, and background
was measured in annuli, with radii of 50$^{\arcsec}$ and 67.2$^{\arcsec}$. 
In some cases, smaller radii were adopted to exclude contributions from nearby sources.
The HR for the IC 2391 members and probable members (top panel in Fig. \ref{hist}, 
see Sect. 2.3) are rather soft, spanning the range from -0.47$\pm$0.14 to 0.15$\pm$0.11 
while for the other X-ray sources (bottom panel in Fig. \ref{hist}) HR spans a wider
range; the vast majority of these X-ray sources have a HR  value higher than that 
of IC 2391 members.

\begin{table*}[!h]
\caption{XMM-Newton sources in the IC 2391 region, as detected in the combined EPIC
 image. The X-ray counts are expressed in "MOS1 equivalent" units.}
\bigskip
\scriptsize
\label{Det}
\begin{tabular}{lllllllllll}
\hline\hline
 \multicolumn{1}{c}{Source} &\multicolumn{1}{c}{Name} &\multicolumn{1}{c}{Ra} &\multicolumn{1}{c}{Dec}
 & \multicolumn{1}{c}{Exposure} &
\multicolumn{1}{c}{Rate $\pm$ Err.} &\multicolumn{1}{c}{Significance} 
&\multicolumn{1}{c}{HR} \\
& \multicolumn{1}{c}{VXR (CTIO)} &\multicolumn{1}{c}{(J2000)} &\multicolumn{1}{c}{(J2000)} & 
\multicolumn{1}{c}{[ks]} & \multicolumn{1}{c}{[cnt/ks]} & & \\
\hline
1 & & 08:40:06.686 & -53:01:33.56 &  18.96 & 6.02 $\pm$ 1.39 & 6.93 &0.95
$\pm$ 0.31 \\
2 & 19 & 08:40:37.416 & -53:07:44.08 &  21.56 & 3.54 $\pm$ 0.74 & 5.69 &-0.40
$\pm$ 0.19  \\
3 & & 08:40:42.883 & -52:58:25.75 &  45.63 & 1.11 $\pm$ 0.28 & 5.74 & &\\
4 & & 08:40:43.087 & -52:49:03.25 &  19.60 & 35.09 $\pm$ 26.16 & 29.86 & -0.40 
$\pm$ 0.12\\
5 &  &08:40:44.654 & -52:58:37.85 &  33.49 & 20.50 $\pm$ 1.41 & 36.12 & 0.24
$\pm$ 0.10\\
6 & & 08:40:47.196 & -53:08:30.91 &  19.74 & 2.26 $\pm$ 0.63 & 5.43 & &\\
7 & & 08:40:49.128 & -52:53:02.98 &  39.78 & 2.57 $\pm$ 0.47 & 7.73 & &\\
8 & & 08:40:49.570 & -52:56:51.94 &  41.28 & 1.51 $\pm$ 0.34 & 5.69 &&\\
9 & & 08:40:49.805 & -53:07:40.80 &  27.50 & 1.92 $\pm$ 0.47 & 5.52 &-0.83
$\pm$ 0.27 \\
10 & & 08:40:51.955 & -53:01:59.59 &  47.06 & 5.26 $\pm$ 0.48 & 20.74 & 0.80
$\pm$ 0.15 \\
11 & & 08:40:55.925 & -52:56:11.58 &  50.32 & 1.89 $\pm$ 0.37 & 6.54 &  &\\
12 & & 08:40:56.256 & -53:03:07.63 &  56.21 & 1.17 $\pm$ 0.25 & 5.84 & &\\
13 & & 08:40:59.563 & -52:57:57.13 &  58.43 & 5.17 $\pm$ 0.47 & 17.30 &0.40
 $\pm$ 0.12\\
14 & & 08:41:02.839 & -52:59:49.78 &  64.05 & 1.00 $\pm$ 0.22 & 5.66 &  &\\
15 & 24A &08:41:04.992 & -52:51:28.69 &  43.48 & 9.81 $\pm$ 0.70 & 23.59 & -0.35$\pm$0.10 \\
16 & 28B & 08:41:07.978 & -53:00:03.82 &  69.10 & 3.95 $\pm$ 0.35 & 19.37 & 0.05 
 $\pm$ 0.09\\
17 & 29& 08:41:09.427 & -53:02:12.98 &  68.87 & 9.01 $\pm$ 0.50 & 34.78 & -0.01
 $\pm$ 0.05\\
18 && 08:41:09.660 & -52:48:39.13 &  30.10 & 3.35 $\pm$ 0.67 & 5.87 & &\\
19 &30& 08:41:10.140 & -52:54:10.55 &  58.37 & 23.82 $\pm$ 0.89 & 51.01 & -0.15
 $\pm$ 0.04\\
20 && 08:41:11.722 & -52:51:24.30 &  47.84 & 3.15 $\pm$ 0.54 & 6.47 &&\\
21 &32C-H (96)& 08:41:12.269 & -53:09:09.90 &  51.11 & 2.82 $\pm$ 0.42 & 9.34 &-0.25
$\pm$ 0.18 \\
22 && 08:41:19.567 & -53:05:29.76 &  70.55 & 1.38 $\pm$ 0.28 & 6.24 &&\\
23 && 08:41:19.970 & -52:58:48.65 &  56.02 & 1.05 $\pm$ 0.29 & 5.69 &&\\
24 && 08:41:21.065 & -53:01:16.36 &  81.78 & 1.37 $\pm$ 0.24 & 7.79&1.00 $\pm$ 0.42\\
25 &33A& 08:41:22.248 & -53:04:46.38 &  75.54 & 3.97 $\pm$ 0.33 & 20.28 & -0.33
$\pm$ 0.09\\
26 && 08:41:22.901 & -52:56:46.93 &  60.85 & 7.23 $\pm$ 0.60 & 24.21 & 0.63
$\pm$ 0.22\\
27 && 08:41:25.481 & -52:57:44.21 &  85.43 & 1.83 $\pm$ 0.26 & 9.43&-0.04
$\pm$ 0.10\\
28 && 08:41:27.254 & -53:02:59.64 &  88.39 & 0.98 $\pm$ 0.16 & 9.26 &0.65
$\pm$ 0.20\\
29 && 08:41:28.474 & -53:13:55.92 &  9.65 & 18.70 $\pm$ 21.07 & 16.50 & &\\
30 && 08:41:31.570 & -52:58:29.21 &  94.72 & 2.94 $\pm$ 0.25 & 20.28
&1.00 $\pm$ 0.20\\
31 & &08:41:33.950 & -53:10:22.37 &  55.84 & 2.43 $\pm$ 0.38 & 8.43 &-0.21
$\pm$ 0.16  \\
32 &(100)& 08:41:35.782 & -53:09:26.53 &  60.82 & 2.70 $\pm$ 0.36 & 12.24 &0.15
$\pm$ 0.11  \\
33 & &08:41:36.178 & -53:06:56.34 &  76.26 & 2.5 $\pm$ 0.27 & 15.14 & 0.90 
$\pm$ 0.21\\
34 &37& 08:41:38.964 & -53:09:24.23 &  62.42 & 22.2 $\pm$ 0.85 & 49.69 & 0.09
$\pm$ 0.04\\
35 &38& 08:41:39.686 & -52:59:33.94 &  105.21 & 79.09 $\pm$ 1.15 & 172.24 & 0.04
$\pm$ 0.02\\
36 && 08:41:40.786 & -53:07:22.22 &  74.24 & 1.97 $\pm$ 0.28 & 9.54  &0.85
$\pm$ 0.23\\
37 && 08:41:41.486 & -53:00:35.03 &  105.09 & 2.10 $\pm$ 0.21 & 15.89 &&\\
38 &(103)& 08:41:43.810 & -53:14:04.49 &  40.85 & 1.52 $\pm$ 0.37 & 5.69 &  &\\
39 && 08:41:45.245 & -52:58:48.79 &  107.08 & 0.78 $\pm$ 0.13 & 9.45 &0.48
$\pm$ 0.16\\
40 && 08:41:45.986 & -53:00:51.84 &  104.14 & 0.56 $\pm$ 0.12 & 5.31 & &\\
41 && 08:41:46.090 & -53:01:45.30 &  108.57 & 0.75 $\pm$ 0.12 & 9.65 & 0.21
$\pm$ 0.17 \\
42 && 08:41:48.979 & -52:55:20.10 &  89.98 & 4.92 $\pm$ 0.32 & 28.72 &0.81
$\pm$ 0.13\\
43 && 08:41:50.400 & -52:57:22.36 &  103.15 & 4.52 $\pm$ 0.28 & 29.53
&1.00 $\pm$ 0.20\\
44 && 08:41:51.418 & -52:49:07.54 &  50.00 & 7.43 $\pm$ 0.59 & 20.03 & -0.06
$\pm$ 0.08\\
45 &40& 08:41:54.137 & -52:57:57.64 &  82.32 & 8.54 $\pm$ 0.54 & 35.87 & -0.35
$\pm$ 0.09\\
46 &41& 08:41:57.898 & -52:52:14.63 &  69.59 & 58.83 $\pm$ 1.23 & 111.92 & -0.06
$\pm$ 0.02\\
47 && 08:41:58.219 & -52:51:32.08 &  65.01 & 1.48 $\pm$ 0.26 & 8.58 &\\
48 && 08:41:58.788 & -52:47:13.16 &  27.20 & 4.73 $\pm$ 1.89 & 11.76 & &\\
49 && 08:41:58.838 & -53:05:02.98 &  98.14 & 0.66 $\pm$ 0.13 & 7.05 &1.00 $\pm$ 0.31\\
50 &(106)& 08:41:59.086 & -53:12:38.52 &  48.86 & 2.00 $\pm$ 0.40 & 6.47 & &\\
51 && 08:41:59.834 & -52:59:53.95 &  113.39 & 1.42 $\pm$ 0.17 & 12.53 &0.76
 $\pm$ 0.17\\
52 && 08:42:00.722 & -53:06:04.57 &  68.53 & 1.29 $\pm$ 0.29 & 5.85 &&\\
53 && 08:42:01.022 & -52:53:01.82 &  73.02 & 1.27 $\pm$ 0.22 & 7.99 &&\\
54 && 08:42:04.346 & -53:09:38.05 &  65.78 & 0.63 $\pm$ 0.17 & 5.28 & &\\
55 &42 (108)& 08:42:05.261 & -52:53:55.36 &  48.28 & 11.00 $\pm$ 0.82 & 27.74 &&\\
56 && 08:42:07.332 & -52:57:52.45 &  107.53 & 2.10 $\pm$ 0.22 & 13.85 & -0.03
$\pm$ 0.09\\
57 && 08:42:07.920 & -53:11:06.25 &  45.73 & 1.27 $\pm$ 0.26 & 7.04 &1.00
$\pm$ 0.17\\
58 && 08:42:08.378 & -53:02:44.81 &  108.9 & 1.22 $\pm$ 0.23 & 5.93 &&\\
59 && 08:42:11.431 & -53:08:01.90 &  75.35 & 2.78 $\pm$ 0.32 & 12.63&0.70
$\pm$ 0.17\\
60 &44+L33& 08:42:12.302 & -53:06:04.68 &  90.68 & 142 $\pm$ 1.70 & 212.75 & -0.04
$\pm$ 0.01\\
61 && 08:42:12.418 & -52:54:16.85 &  83.38 & 0.82 $\pm$ 0.17 & 5.86 &&\\
62 && 08:42:13.102 & -52:57:50.04 &  105.31 & 0.57 $\pm$ 0.11 & 7.41 &&\\
63 &45 &08:42:14.779 & -52:56:02.40 &  94.06 & 143.35 $\pm$ 1.63 & 226.53 &0.01
$\pm$ 0.01\\
64 &47& 08:42:18.514 & -53:01:58.51 &  107.73 & 51.98 $\pm$ 0.93 & 138.66 & -0.07
$\pm$ 0.02\\
65 & &08:42:18.898 & -52:51:54.65 &  64.79 & 2.14 $\pm$ 0.28 & 11.23&&\\
66 &46& 08:42:19.068 & -53:06:00.68 &  88.64 & 11.50 $\pm$ 0.50 & 47.78 & -0.07
$\pm$0.05\\
67 && 08:42:19.687 & -52:50:09.71 &  55.14 & 1.70 $\pm$ 0.28 & 8.48 &-0.59
$\pm$ 0.22 &\\
68 && 08:42:22.066 & -53:04:21.76 &  96.30 & 1.94 $\pm$ 0.22 & 14.37 &  &\\
69 && 08:42:23.047 & -52:48:52.42 & 35.06 & 1.43 $\pm$ 0.40 & 6.21 & &\\
70 && 08:42:23.652 & -53:10:25.75 &  58.28 & 1.25 $\pm$ 0.26 & 6.28&&\\
71 && 08:42:23.854 & -53:01:46.70 &  103.46 & 1.95 $\pm$ 0.21 & 14.79&&\\
72 && 08:42:28.294 & -52:51:18.58 &  58.08 & 2.68 $\pm$ 0.43 & 7.42 &0.47
$\pm$ 0.22\\
73 && 08:42:30.564 & -53:03:40.36 &  63.19 & 2.36 $\pm$ 0.73 & 10.72 & &\\
74 &49& 08:42:30.811 & -52:57:31.50 &  91.91 & 28.01 $\pm$ 0.74 & 81.05 &-0.09
$\pm$ 0.03\\
75 && 08:42:37.176 & -52:49:50.09 &  38.30 & 1.01 $\pm$ 0.35 & 5.70 & &\\
76 && 08:42:37.349 & -52:57:26.24 &  85.90 & 8.49 $\pm$ 0.43 & 38.53 & 0.74
$\pm$ 0.09\\
77 && 08:42:37.819 & -52:56:25.04 &  81.26 & 0.71 $\pm$ 0.15 & 6.43 & &\\
78 && 08:42:38.311 & -53:03:29.92 &  77.39 & 1.63 $\pm$ 0.31 & 6.06 &&\\
79 && 08:42:39.595 & -53:05:39.30 &  76.26 & 4.79 $\pm$ 0.36 & 23.98 & 0.22
$\pm$ 0.07\\
80 && 08:42:44.196 & -53:10:41.09 &  49.14 & 2.69 $\pm$ 0.52 & 6.30 &0.05
$\pm$ 0.24\\
\hline
\end{tabular}
\end{table*}
\begin{table*}[!h]
\addtocounter{table}{-1}
\caption{Continued}
\bigskip
\scriptsize 
\label{Det}
\begin{tabular}{lllllllllll}
\hline\hline
 \multicolumn{1}{c}{Source}& \multicolumn{1}{c}{Name} &\multicolumn{1}{c}{Ra} &\multicolumn{1}{c}{Dec}
 & \multicolumn{1}{c}{Exposure} &
\multicolumn{1}{c}{Rate $\pm$ Err.} &\multicolumn{1}{c}{Significance} 
&\multicolumn{1}{c}{HR} \\
& \multicolumn{1}{c}{VXR} &\multicolumn{1}{c}{(J2000)} &\multicolumn{1}{c}{(J2000)} & 
\multicolumn{1}{c}{[ks]} & \multicolumn{1}{c}{[cnt/ks]} & &  \\
\hline
81 && 08:42:45.293 & -52:55:15.89 &  69.40 & 3.69 $\pm$ 0.42 & 11.02 & 0.58
$\pm$ 0.17\\
82 &52& 08:42:46.601 & -53:01:01.96 &  81.31 & 67.59 $\pm$ 1.24 & 128.9 &-0.03
$\pm$  0.02\\
83 && 08:42:47.004 & -52:56:24.40 &  73.08 & 0.99 $\pm$ 0.21 & 6.68 &&\\
84 && 08:42:47.664 & -53:09:22.39 &  54.97 & 1.47 $\pm$ 0.29 & 7.25 & &\\
85 && 08:42:48.245 & -53:03:21.96 &  75.03 & 2.85 $\pm$ 0.39 & 9.10 &-0.29
$\pm$ 0.12\\
86 &53 (126)& 08:42:49.085 & -52:52:17.65 &  55.14 & 4.27 $\pm$ 0.45 & 13.5 &-0.47
 $\pm$ 0.14\\
87 && 08:42:49.565 & -53:07:02.42 &  61.99 & 1.18 $\pm$ 0.27 & 5.36 &&\\
88 && 08:42:50.690 & -52:49:57.94 &  19.78 & 3.27 $\pm$ 0.74 & 7.12 &&\\
89 && 08:42:54.653 & -53:00:49.61 &  73.31 & 1.49 $\pm$ 0.27 & 6.75 &&\\
90 && 08:42:55.807 & -52:55:22.30 &  56.28 & 3.8 $\pm$ 0.38 & 17.13 &
1.00 $\pm$ 0.23\\
91 && 08:42:56.861 & -52:51:09.25 & 34.51 & 1.25 $\pm$ 0.37 & 6.01 &&\\
92 &55& 08:42:57.497 & -53:08:54.56 &  40.51 & 3.5 $\pm$ 0.70 & 9.59 &&\\
93 && 08:42:58.764 & -53:05:55.03 &  61.41 & 6.89 $\pm$ 0.49 & 25.14 & -0.13
$\pm$ 0.06\\
94 && 08:43:03.564 & -52:56:13.67 &  36.15 & 2.36 $\pm$ 0.53 & 6.76& &\\
95 &56& 08:43:03.571 & -53:04:41.74 &  53.81 & 50.91 $\pm$ 1.37 & 82.85 & -0.03
$\pm$ 0.03\\
96 && 08:43:07.279 & -53:01:51.35 &  57.99 & 1.36 $\pm$ 0.27 & 7.25 &-0.07
$\pm$ 0.14 \\
97 && 08:43:11.659 & -53:04:13.33 &  44.89 & 2.41 $\pm$ 0.4 & 10.03 &0.30
 $\pm$ 0.18\\
98 && 08:43:13.826 & -52:56:19.18 &  49.72 & 2.15 $\pm$ 0.40 & 7.34 &0.50
$\pm$ 0.24\\
99 &&08:43:19.913 & -53:05:34.33 &  20.34 & 3.57 $\pm$ 0.75 & 7.37 & &\\
\hline
\\
\\
\end{tabular}
\end{table*}
\normalsize

\begin{table*}[!h]
\caption{X-ray and optical properties of IC 2391 members, and upper limits
 in the combined EPIC image. Asterisks indicate the members
for which spectral and timing analysis has been done using the EPIC/pn camera. "?" = suspected cluster member; "\#" = observed only by MOS1 and MOS2.}
\bigskip
\scriptsize
\begin{tabular}{lllllllllllllll}
\hline
\hline
\multicolumn{1}{l}{Name} &\multicolumn{1}{l}{Flag} &\multicolumn{1}{l}{Source} &\multicolumn{1}{c}{Ra$_{ott}$}&\multicolumn{1}{c}{Dec$_{ott}$} &\multicolumn{1}{c}{V} & \multicolumn{1}{c}{B-V}& \multicolumn{1}{c}{R-I} &\multicolumn{1}{l}{V-I} &\multicolumn{1}{c}{Rate}&\multicolumn{1}{c}{K-S}
&\multicolumn{1}{c}{logL$_x$}  \\
&& &\multicolumn{1}{c}{(J2000)} &\multicolumn{1}{c}{(J2000)} &\multicolumn{1}{c}{(I)} & & &
&\multicolumn{1}{c}{[cnt/ks]} & & \multicolumn{1}{c}{[erg/s]} &  &  \\
\hline
L06 && & 08:40:17.501 & -53:00:55.69 & 5.57 & -0.13 &    &    & $<$2.11 &  &$<$28.44  \\
VXR19& ? &2& 08:40:36.800 & -53:07:51.00 & 17.01 &    & 1.56 & 2.77 & 3.54 &$<$90\% &
28.66\\
VXR24A&? &15& 08:41:04.100 & -52:51:44.00 & 15.85 &    & 1.45 & 2.58 & 9.81 &$90\% - 95\%$ &29.10 \\
VXR28B&? &16& 08:41:08.000 & -53:00:08.00 & 17.75 &    & 1.65 & 2.94 & 3.95 &$\geq99\%$  &28.71\\
VXR29&? &17& 08:41:09.400 & -53:02:1:00   & 17.40 &    & 1.71 & 3.07 & 9.01 &$\geq99\%$  &29.07\\
VXR30&* &19& 08:41:09.998 & -52:54:11.24 & 9.83 & 0.49 &    &    & 23.82 &$<90\%$ & 29.49 \\
VXR32C$^a$&? &21& 08:41:11.400 & -53:09:16.00 & 15.34 &    & 0.99 & 2.05 & 2.82 &$\geq99\%$  &28.08\\
VXR32H$^a$&? &21& 08:41:11.400 & -53:09:16.00 & 19.64 &    & 1.91 & 3.50 & 2.82 &$\geq99\%$  &28.08\\
CTIO 096$^a$& &21& 08:41:12.370 & -53:09:10.30 &(16.14)  &   &1.91   &   & 2.82 &$\geq99\%$  &28.08\\
H21& && 08:41:16.314 & -53:04:26.66 & 11.69 & 0.82 &    &    & $<$0.76 & &$<$27.99\\
VXR33A&? &25& 08:41:22.300 & -53:04:55.00 & 16.93 &    & 1.64 & 2.91 & 3.97 &$\geq99\%$ & 28.71\\
CTIO 100$^b$& &32& 08:41:35.960 & -53:09:27.10 &18.46  &   &1.80   &   & 2.70 &$<90\%$ &
28.54\\
VXR37&*? &34& 08:41:38.500 & -53:09:35.00 & 17.10 &    & 1.66 & 2.87 & 22.20 &$\geq99\%$&29.46 \\
VXR38&* &35& 08:41:39.800 & -52:59:36.00 & 13.38 & 1.25 & 0.77 & 1.54 & 79.09 &$\geq99\%$  &30.01\\
CTIO 103& &38& 08:41:43.960 & -53:14:07.00 &17.74  &   &1.72   &   & 1.52 & $95\% - 99\%$ &28.29  \\
L15& && 08:41:46.574 & -53:03:45.37 & 7.59 & 0.27 &    &    & $<$0.52 & &27.83 \\
SHJM7$^a$& && 08:41:52.000 & -53:06:47.00 & 12.52 & 1.10 & 0.58 & 1.21 &$<$0.69 & &$<$27.95 \\
VXR40&? &45& 08:41:53.700 & -52:58:07.00 & 17.14 &    & 1.62 & 2.86 & 8.54  &$<90\%$
&29.04\\
VXR41&* &46& 08:41:57.900 & -52:52:16.00 & 13.48 & 1.25 & 0.72 & 1.46 & 58.83 &$<90\%$  &29.88\\
CTIO 106& &50& 08:41:58.930 & -53:12:36.30 &(16.45)  &   &1.98   &   & 2.00 &$<90\%$  &27.93 \\
CTIO 106& &50& 08:41:58.930 & -53:12:36.30 &(16.45)  &   &1.98   &   & 2.00 &$<90\%$  & 27.93\\
CTIO 106& &50& 08:41:58.990 & -53:12:36.90 &(16.56)  &   &2.01   &   & 2.00 &$<90\%$  & 27.93 \\
CTIO 108& &55& 08:42:04.910 & -52:53:54.10 &(13.32)  &   & 1.41  &   & 11.00 &$90\% - 95\% $  &28.85 \\
VXR42&? &55& 08:42:05.300 & -52:54:00.00 & 15.88 &    & 1.37 & 2.49 & 11.00 &$90\% - 95\%$  &28.85 \\
L24& && 08:42:07.984 & -53:09:35.15 & 8.60 & 0.15 &    &    & $<$0.84 &  & $<$28.04\\
L13& && 08:42:09.955 & -52:58:03.99 & 7.39 & 0.02 &    &    & $<$0.55 &  &
$<$27.85\\
VXR44&*? &60& 08:42:12.000 & -53:06:12.00 & 9.69 & 0.42 &    &    & 142.00 &$\geq99\%$  &29.96\\
L33& &60& 08:42:12.331 & -53:06:04.50 & 9.59 & 0.44 &    &    & 142.00
&$\geq99\%$  & 29.96\\
VXR45&*? &63& 08:42:14.500 & -52:56:12.00 & 10.70 & 0.81 & 0.44 & 0.90 & 143.35 &$95\%-99\%$ &30.27  \\
VXR46&* &66& 08:42:19.027 & -53:06:00.45 & 5.52 & -0.15 &    &    & 11.50 &$95\%-99\%$ & 29.17\\
VXR47&* &64& 08:42:18.400 & -53:01:57.00 & 14.00 & 1.44 & 1.10 & 2.03 & 51.98 &$\geq99\%$ & 29.72\\
CTIO 114$^a$& && 08:42:21.560 & -52:53:38.90 &(12.51)  &   &0.85   &   & $<$0.68  & &
$<$27.94\\
VXR48& && 08:42:25.426 & -53:06:50.49 & 4.84 & -0.17 &    &    & $<$0.75  & &
$<$27.99\\
VXR49B$^b$&*? &74& 08:42:29.700 & -52:57:43.00 & 14.34 &    & 0.96 & 1.89 & 28.01 &$\geq99\%$ &29.56 \\
VXR52&* &82 &08:42:46.640 & -53:01:01.84 & 10.34 & 0.57 & 0.23 & 0.44 & 67.59 &$\geq99\%$ &29.94 \\
VXR53&? &86& 08:42:48.700 & -52:52:18.00 & 17.61 &    & 1.79 & 3.22 & 4.27 & $<90\%$&
28.44 \\
CTIO 126& &86& 08:42:49.050 & -52:52:15.90 & (14.39) &   &1.80   &   & 4.27  &$<90\%$ &28.44\\
H36& && 08:42:50.108 & -53:02:01.96 & 11.31 & 0.64 &    &    & $<$0.84 &  &
$<$28.04\\
VXR55&? & 92& 08:42:57.900 & -53:06:04.00 & 16.29 &    & 1.58 & 2.82 & 6.89 &$90\%-95\%$ & 28.95 \\
CTIO 130$^b$&\# && 08:42:58.210 & -52:49:46.20 &(17.23)  &   &2.03   &   & $<$1.99  &
&$<$28.41 \\
VXR56&* &95& 08:43:03.464 & -53:04:41.36 & 7.68 & 0.10 &    &    & 50.91  &$>99\%$ &
29.82\\
CTIO 136& && 08:43:15.140 & -52:58:23.00 & 19.72&   &1.98   & & $<$1.08  & &
$<$28.15 \\
\hline
\end{tabular}
$^a$ These stars have R-I inconsistent with cluster membership (see Fig. \ref{LR}).\\
$^b$ Stars having (J-K), J values incompatible with the cluster main sequence.
\label{mem}
\end{table*}
\normalsize

\subsection{Cluster members and identifications}
We have collected an optical catalog of cluster members based on 
the Prosser \& Stauffer Open Cluster Database project 
\footnote{available at http://www.noao.edu/noao/staff/cprosser}, and the 
\citet{Ba01} catalog for the faintest stars.
Our final list contains 172 likely members with spectral types ranging from B to M. \\
The EPIC field of view, shown in Fig. \ref{dss}, contains 42 out of the 172  
members of the cluster with spectral type from B to M.
We have  cross-identified the X-ray sources with the optical catalog, 
in  two steps; in the first step we found a systematic offset between X-ray 
source positions and membership catalog of 1.02${\arcsec}$ in RA and 1.23${\arcsec}$ 
in Dec; in the second step, we corrected the X-ray positions  for this systematic 
offset, before matching the X-ray and optical member positions, and retained
an identification if the offset between X-ray and optical positions was less 
than 14${\arcsec}$.
The choice of such  a limiting distance is a good compromise between the attempt
to minimize the number of spurious identifications (whose expected number for an offset 
$\sim$ 14${\arcsec}$ is 0.76)  and to keep the largest number of bona-fide 
optical  counterparts.

We have found  24 X-ray sources identified with 31 possible or probable  members;  
in particular we have detected about 30\%  of type B and A, 100\% of F, 87\% of G-K 
and 87\% of M member stars in the FOV.
In five cases it has been impossible to resolve very close stars, owing to the
limited spatial resolution  of the X-ray telescope. 
For these unresolved sources, the X-ray flux has been divided evenly 
between the optical candidates in the absence of further information. 

We have also obtained upper limits to count rates at the 
optical positions of undetected cluster members. 
Table \ref{mem} summarizes the optical and X-ray characteristics of 
the possible and probable  members of the cluster. 

We have also searched optical counterparts of the 75 X-ray sources not 
identified with known members of the cluster. 
Using the 2MASS All-Sky Catalog of Point Sources \citep{Cut03} with a 
match radius of 14${\arcsec}$, we found that 64 out of 75 X-ray sources 
have counterparts in the 2MASS catalog.
Among these 64 sources, 9 have infrared photometry consistent with the cluster
sequence (see Table \ref{JHK}); 
for 4 out of these 9 sources it has been possible to compute a reliable HR
and in 3 case the values indicate a quite soft spectrum, with an HR in the same range
as that of the IC 2391 members.
The positions of the stars CTIO 100, CTIO 130 and VXR49B (which have been 
reported as members or probable members in the IC 2391 optical catalogues)  
in the (J-K), J color-magnitude diagram, are inconsistent with the cluster 
main sequence, making their membership doubtful.

\begin{figure*}[t]
\centering
\centerline{\psfig{figure=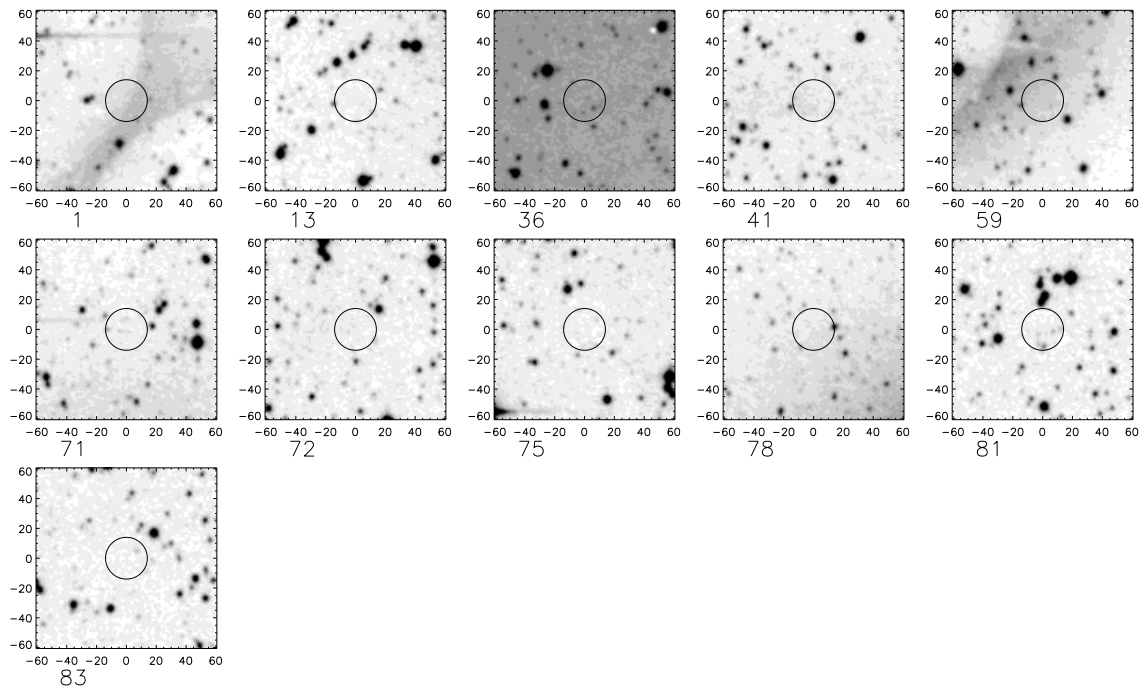,width=16cm}}
\caption{Finding charts for unidentified EPIC sources in the IC 2391. 
Each finding chart is a square of 2${\arcmin}$ of size centered on X-ray positions; 
we also show the 14${\arcsec}$ acceptance  circle for X-ray/optical match.}
\label{char}
\end{figure*}

Eleven X-ray sources have no catalogued counterpart. A search on
the STScI Digitised Sky Survey (DSS) shows that most of these sources have faint
counterparts.
Fig. \ref{char} shows the IR DSS finding charts for these 11 sources. 

\begin{table*}
\caption{Infrared photometry of the nine counterparts of X-ray sources whose 2MASS
photometry is consistent with the IC 2391 main sequence."Y" indicates the X-ray sources
having the HR in the same range as IC 2391; "N"= the X-ray source with the HR outiside the IC 2391 member range.}
\bigskip
\scriptsize
\centering
\begin{tabular}{llllll}
\hline
\hline
\multicolumn{1}{l}{Source}& \multicolumn{1}{c}{Flag} & \multicolumn{1}{c}{J} & \multicolumn{1}{c}{H}& \multicolumn{1}{c}{K} \\
\hline
11&&13.039 &12.349&12.157\\
18&&13.524 &12.904&12.646\\
20&&14.955 &14.344&13.987\\
23&&14.143 &13.583&13.235\\
39&N&14.767&13.999&13.845\\ 
44&Y&13.717&13.128&12.869\\
85&Y&13.587&12.989 &12.677\\
93&Y&12.010&11.384&11.128\\
99&&14.560&13.901&13.612 \\
\hline
\label{JHK}
\end{tabular}
\end{table*}
\normalsize

\subsection{Analysis of IC 2391  X-ray sources}
For the eleven  X-ray sources of the cluster with sufficient signal to noise 
(i.e. more than 500 counts in the EPIC/pn detection), we have performed 
spectral analysis by using primarily data from EPIC/pn, which collected 
typically a factor $\sim$  1.9 more counts-per-source than a single MOS, 
thanks to its higher sensitivity. 
Source and background regions were  selected interactively  using the
Astronomical Data Visualization DS9 display software, with the background regions 
extracted on the same chip and at the same off-axis angle of the source region. 
We have taken into account the shape of the Point Spread Function (PSF) 
to choose the radius of the circle within which to collect source counts. 
The shape of the PSF is quite complex but the radially averaged profile can be 
suitably represented by an analytical function described by \citet{Ghi01} 
and \citet{Saxon02}. Given that the encircled energy fraction is only weakly 
dependent on the off-axis angle we have adopted a constant extraction radius 
of 45${\arcsec}$ for all sources, and corrected for the lost photon counts 
with the proper fraction derived from the $PSF$.
In some cases we have used a  smaller radius to exclude contributions from 
nearby stars and have changed the $PSF$ correction accordingly. 

\section{Spectral analysis }
For the eleven sources with enough count statistics we have extracted spectra 
of the single and double events with energy in the 0.3 - 3.5 keV  band. 
We have adopted this selection, since at higher energies the background 
contribution is dominant while the source counts are very few. 
We have computed the photon redistribution matrix using the SAS task  
{\em rmfgen} and the ancillary response files using the SAS task {\em arfgen}. 
Finally, spectra have been rebinned in order to have at least 25 counts per 
spectral bin.\\ 
We have performed a spectral analysis based on global fitting, using the APEC 
model spectra \citep{Smi01} implemented in the XSPEC software, in order to 
obtain the relevant temperature components and relative Emission Measure (EM)
values.   

The X-ray emission from stellar coronae is expected to be a superposition of 
spectra of many individual loops of hot plasma at different temperatures. 
This implies that the form of the differential emission measure as a function 
of  temperature may be complex. 
Furthermore, chemical abundances may differ from the solar one. 
We have approximated the differential emission measure with one or more 
(in fact at most two) isothermal components and estimated the abundances 
parameter, Z, in which the abundances of individual elements are fixed to the 
ratios observed in the solar photosphere.
Only in one case (VXR52) did we find that a model allowing abundances of Fe,
O, and Mg to vary is needed to improve the fit (see Table \ref{sp}).  
We show the spectra of the brightest stars in Fig. \ref{spe} and report the 
best-fit parameters  in Table \ref{sp}.

In all cases the hydrogen column density N$_H$ is consistent with the photometric 
derived value, 3.2 $\times$ 10$^{19}$ cm$^{-2}$, computed from the average B-V color
excess E$_{B-V}$=0.006 $\pm$ 0.005 \citep{Pa96}. 
The 2-T model fits  result in the emission measure ratios, EM2/EM1, being 
greater than 1,  indicating the presence of significant hot components.\\
We note an increase of the coronal temperature from F to M stars,
as shown in Table \ref{sp}.
In the Pleiades stars, using PSPC data, \citet{Gag95} have found a dependence 
of fitted coronal temperature on spectral type and rotation, with fast G 
rotators hotter than the other late-type stars, and F stars softer than all 
the others. We have not enough spectra to test if this is also the case for IC 2391.
\citet{Brig03}, analysing XMM-Newton data, found an  increase of temperature 
with spectral type (from F to K) in the  Pleiades; this is similar to the trend we see 
in IC 2391. 

Our spectra are systematically fitted  with "depleted" abundances with respect 
to the solar ones. In particular, we found Z $\sim$ 0.1-0.2 Z$_{\odot}$ 
for K-stars and M-stars, $\sim$ 0.3 Z$_{\odot}$ for the G-star VXR45  
and $\sim$ 0.5 Z$_{\odot}$ for the F-stars.
Indications of subsolar abundances with a similar trend of Z with spectral type 
have been found for the Pleiades by \citet{Brig03}.

\begin{table*}
\caption{Results of global fits of EPIC/PN spectra of the X-ray brightest cluster sources.
All fits assume N$_H$ fixed at 3.2 $\times$ 10$^{19}$ cm$^{-2}$.  The 1-T or 2-T in the model, 
as well as the corresponding emission measures and the abundance, Z, were allowed to vary as free parameters in the fitting process. Columns show: (1) Name of the star, (2) spectral type, (3) cool temperature, (4) hot 
temperature, where required, (5) abundances, (6) and (7) logarithm of the  emission measure of the cool and hot 
component respectively, (8) $\chi^2$ / degrees of freedom and null hypothesis probability. The error ranges refer to 90\% confidence intervals.} 
\bigskip
\tabcolsep 0.1truecm
\scriptsize 
\label{sp}
\begin{tabular}{llllllllllll}
\hline
\hline
\multicolumn{1}{l}{Name} &\multicolumn{1}{c}{Source}&\multicolumn{1}{c}{Sp.} &\multicolumn{1}{c}{PN} &\multicolumn{1}{c}{kT$_1$}&\multicolumn{1}{c}{kT$_2$} &\multicolumn{1}{c}{Z/Z$_{\odot}$}&\multicolumn{1}{c}{log EM$_{1}$}&\multicolumn{1}{c}{log EM$_{2} $ } & \multicolumn{1}{c}{$\chi^2$ (dof) P($\chi>\chi_0)$}  \\ 
\multicolumn{1}{c}{VXR} & &\multicolumn{1}{c}{type}&\multicolumn{1}{c}{Counts} &\multicolumn{1}{c}{ [keV] }&\multicolumn{1}{c}{[keV]} &\multicolumn{1}{c}{}&\multicolumn{1}{c}{[cm$^{-3}$]}&\multicolumn{1}{c}{[cm$^{-3}$]} & \\
\hline
46 &66& B8.5IV &601  &  0.69 (0.61-0.76) &  & 0.10 (0.07-0.16) & 52.46 (52.35-52.55) &  & 0.93 (23)~  0.55\\
& &&&&&&\\
56 &95& A1IV &1329&  0.80 (0.74-0.85) & &  0.10 (0.07-0.12) & 52.94 (52.89-52.99) &   &0.87(49)~   0.72\\
& &&&&&&\\
30 &19& F6V &566& 0.60 (0.56-0.64)&   &  0.57 (0.52-0.63) & 52.18 (52.14-52.21) & &0.96 (31)~  0.55\\
& &&&&&&\\
44+L33&60& F6V&5911 &0.65 (0.63-0.66) &  &0.52 (0.44-0.62)  & 52.98 (52.91-53.03) &  &  1.44 (130)~  8.17e-4\\
& &&&&&&\\
52 &82 &F9 &2249&0.59 (0.56-0.61) &  &0.38$^a$ (0.32-0.46)   &  52.72 (52.67-52.77) &    & 1.06 (63)~   0.35\\
& &&&&&&\\
45 &63& G9 &5794  &  0.52 (0.46-0.61) &0.99 (0.94-1.03) &0.27 (0.23-0.34) &52.79  (52.71-52.85) &52.96 (52.91-52.99) & 0.96 (156) ~ 0.62\\
& &&&&&&\\
38 & 35 & K7.5e &2796& 0.40 (0.34-0.66) & 1.21 (1.06-1.27) & 0.18 (0.10-0.19)&52.47 (52.33- 52.56) & 52.77 (52.73-52.78)  &  0.96 (90) ~  0.59\\
& &&&&&&\\
41 &46& K7.5e&1800& 0.32 (0.27-0.38) & 1.01 (0.95-1.07)& 0.22 (0.17-0.29)&  52.51 (52.47-52.57) & 52.78 (52.75-52.80) &  1.03(56) ~  0.42\\
& &&&&&&\\
37 &34&  M &652& 1.00 (0.81-1.05) && 0.10 (0.06-0.15) &52.67 (52.64-52.76) &    &  1.09 (19) ~  0.36\\
& &&&&&&\\
49 &74 & M1e &923& 0.75 (0.69-0.80) &  &  0.11 (0.08-0.15) & 52.59 (52.52-52.66) &    &  1.17 (25)~   0.25\\
& &&&&&&\\
47 &64& M2e &2244& 0.37 (0.33-0.41) & 1.17 (1.04-1.26) &  0.20 (0.13-0.30) & 52.57 (52.53-52.60) & 52.60 (52.52-52.63) & 1.03 (74)~  0.41\\
& &&&&&&\\
\hline
\end{tabular}
\\ $^a$: for VXR52 the model is 1-T VAPEC, Z/Z$_{\odot}$ refers to Fe abundance, 
Mg abundance is 0.28 (0.02-0.54), O abundance is 0.15 (0.05-0.26).\\
\end{table*}
\normalsize

\begin{figure*}[!h]
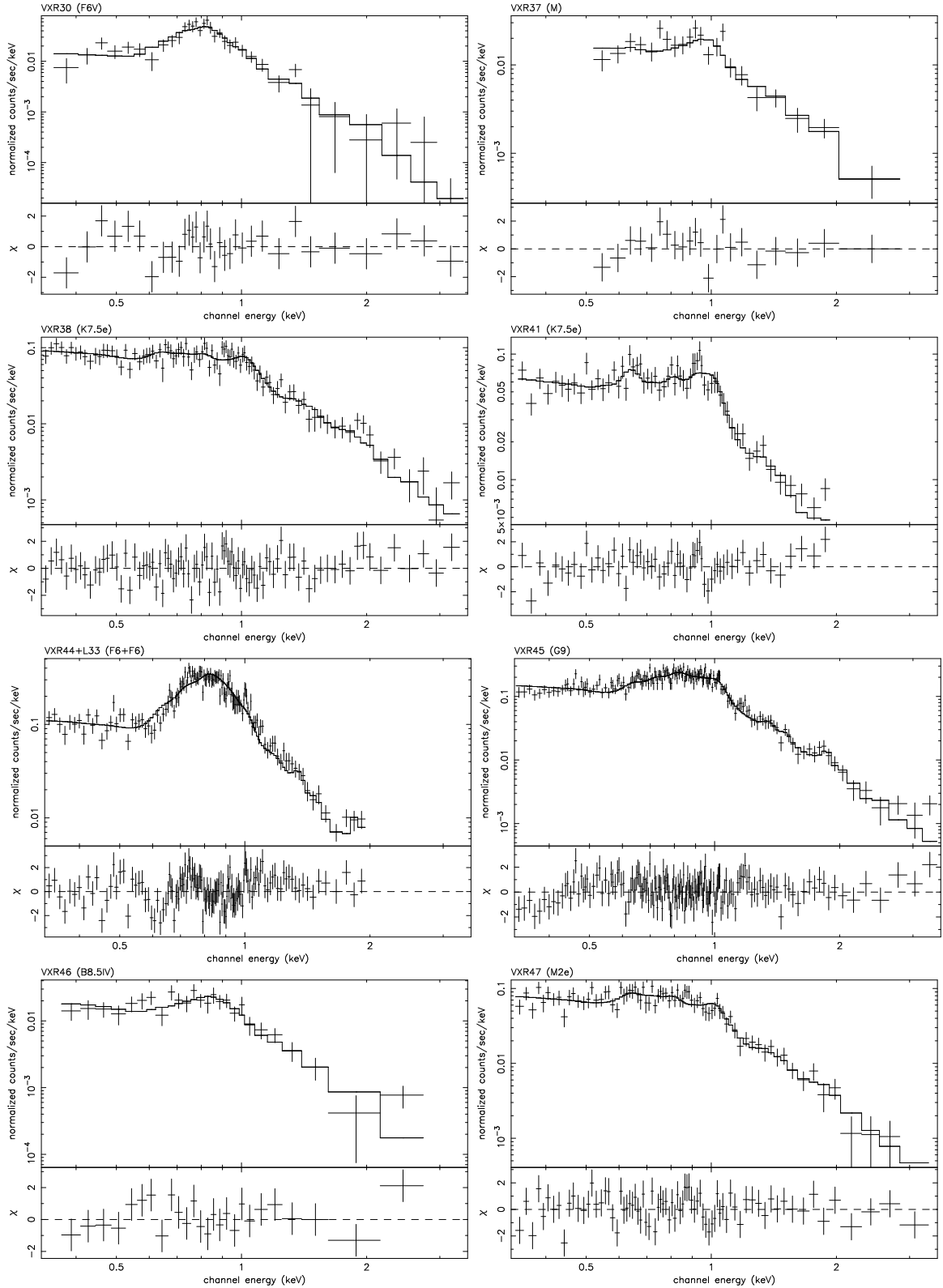

\centerline{\psfig{figure=1525f5a.ps,width=7.6cm,angle=-90} \psfig{figure=1525f5b.ps,width=7.6cm,angle=-90}}
\centerline{\psfig{figure=1525f5c.ps,width=7.6cm,angle=-90} \psfig{figure=1525f5d.ps,width=7.6cm,angle=-90}}
\centerline{ \psfig{figure=1525f5e.ps,width=7.6cm,angle=-90} \psfig{figure=1525f5f.ps,width=7.6cm,angle=-90}}
\centerline{\psfig{figure=1525f5g.ps,width=7.6cm,angle=-90} \psfig{figure=1525f5h.ps,width=7.6cm,angle=-90}}
\caption{Spectra of X-ray bright sources of IC 2391 in the XMM-Newton EPIC field. 
We show the model described in Sect. 3. and in Table \ref{sp}. The lower panel of each
spectrum shows the $\chi^2$ residuals.}
\label{spe}
\end{figure*}
\addtocounter{figure}{-1}
\begin{figure*}[t]
\centerline{\psfig{figure=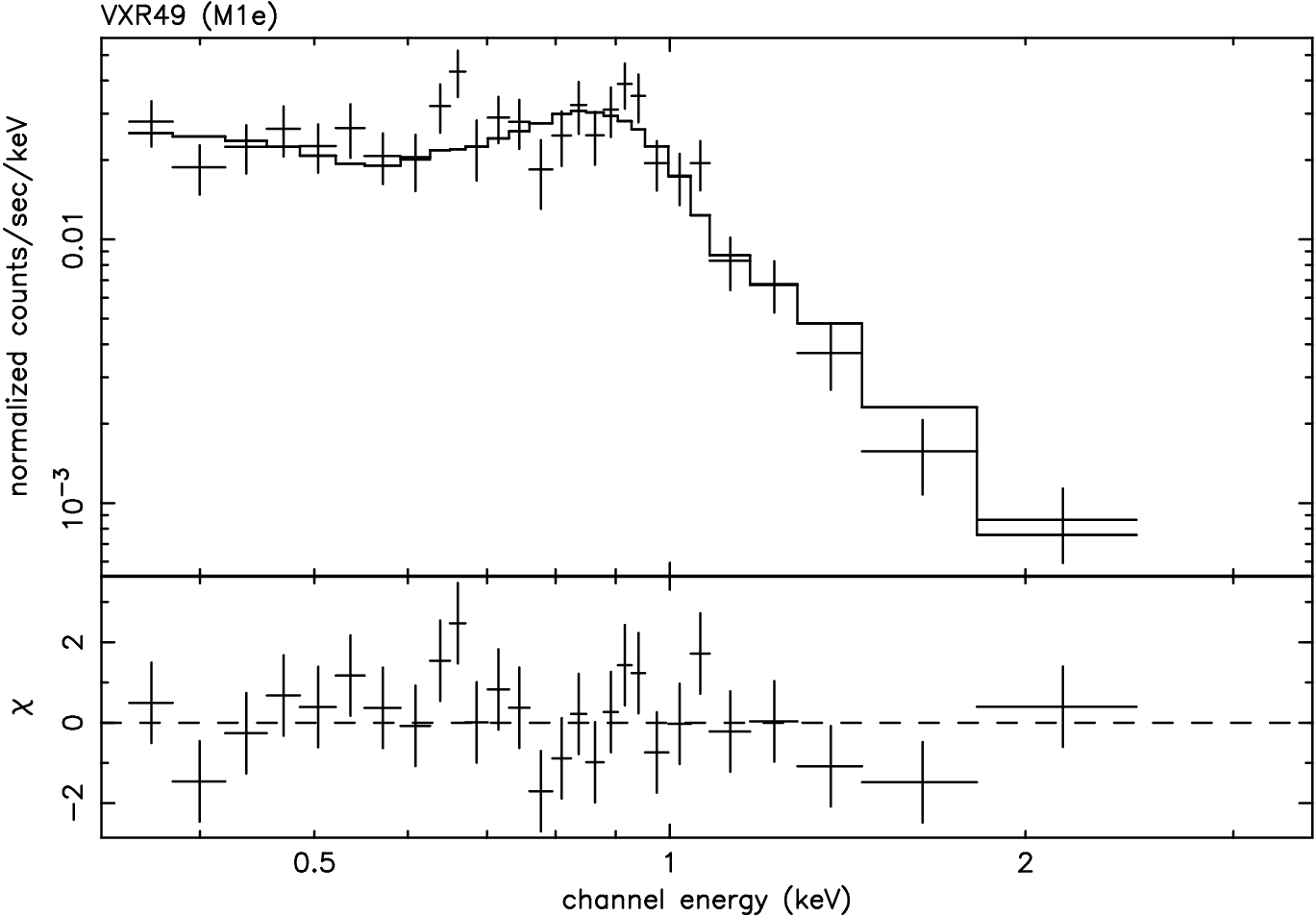,width=7.6cm,angle=-90} \psfig{figure=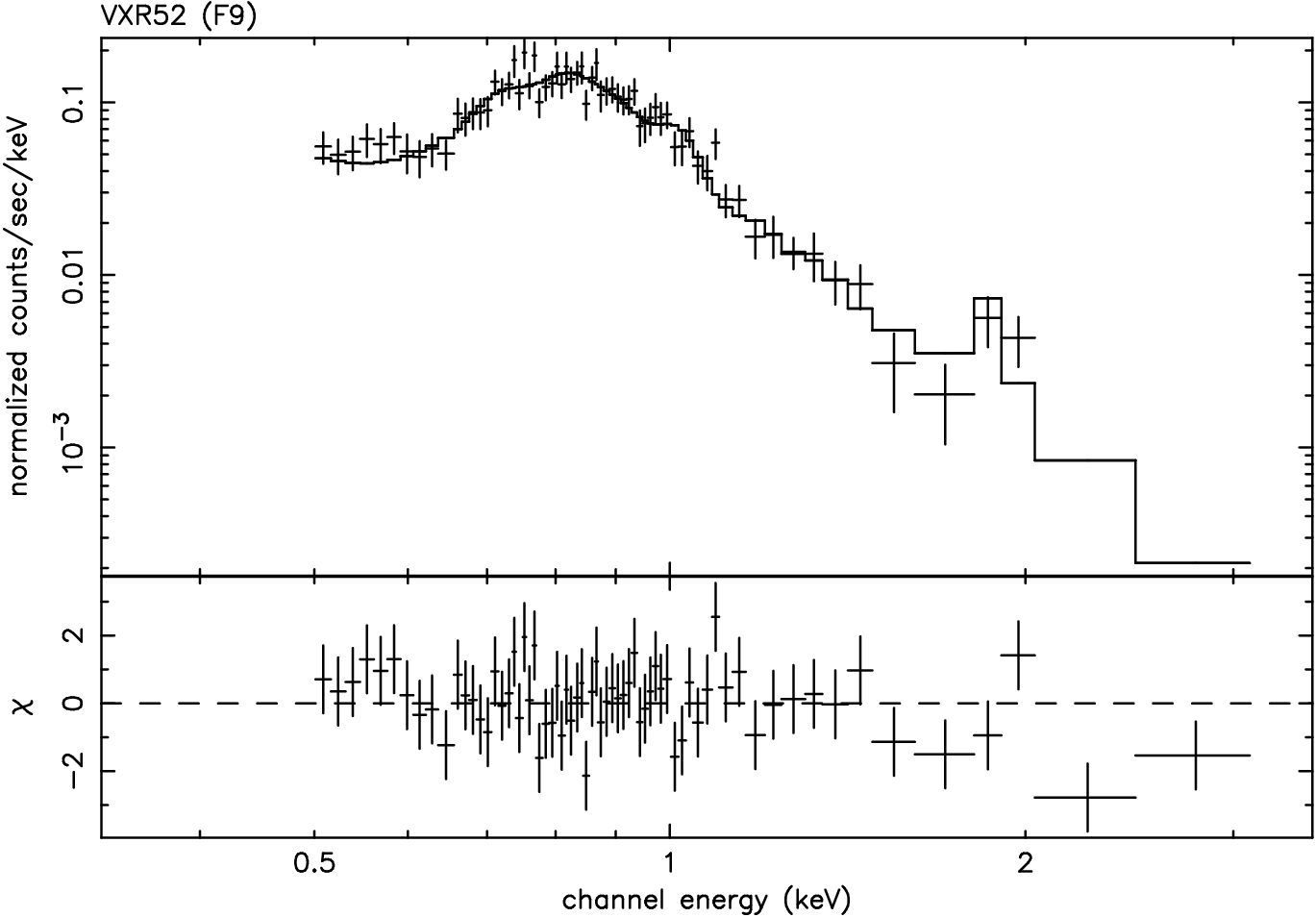,width=7.6cm,angle=-90}}
\centerline{\psfig{figure=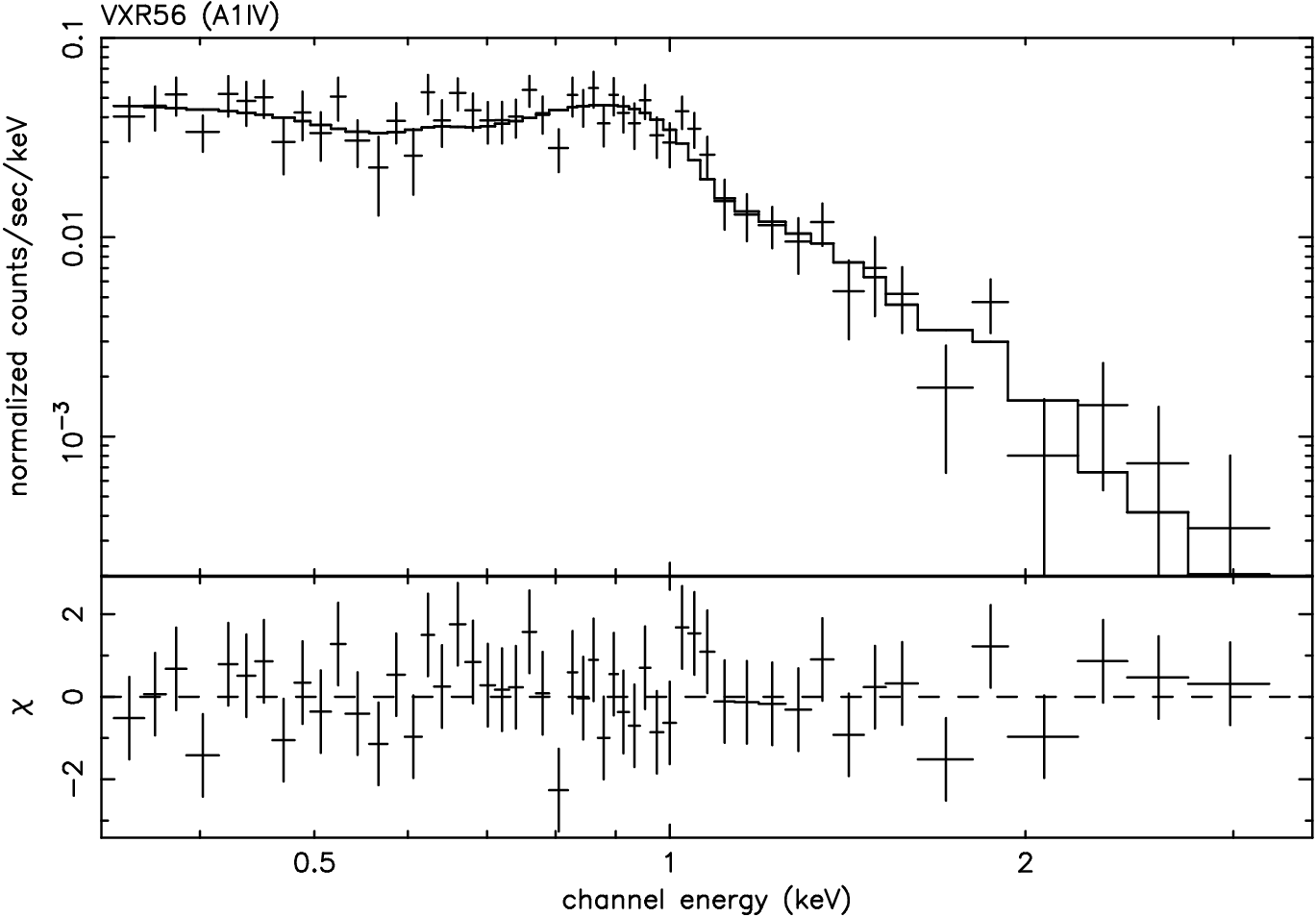,width=7.6cm,angle=-90}}
\label{spe}
\caption{Continued.}
\label{spe}
\end{figure*}

In general, although the obtained fits are statistically
acceptable, some spectral features are not well reproduced. 
An inspection of the spectra (Fig. \ref{spe}) shows that some line
features (e.g. at $\sim$ 0.6 keV and $\sim$ 1 keV) appear to be 
not well reproduced by model fits.
Furthermore, we have been  unable to fit the data of the two F6 
stars, VXR44 and L33, unresolved in the EPIC image. 
The attempts to fit this spectrum by adding further temperature 
components, fixing  the global chemical abundances and hydrogen 
column density or allowing them to vary, did not improve the fit.\\
The stellar structure models predict that stars earlier than A7  
lack of, or have  very thin, convective zones (required to generate 
magnetic activity), and hence lack one key ingredient for an  
$\alpha$-$\Omega$ dynamo. Therefore one expects no X-ray emission 
from these stars. For the binary systems unresolved in the X-ray band 
the emission is usually attributed to late companions. 
This could be the case for two stars of IC 2391: VXR46,
a B8.5 spectral type star flagged as single star by \citet{Pa96}, but that 
could have a faint  unknown companion,  and VXR56, an A1 star flagged as 
spectroscopic binary \citep{Pa96}. 
The X-ray spectra of both  VXR46 and VXR56 are fitted with a 1-T, low Z model, 
quite similar to the best-fit  model for the X-ray spectra of M-type members, 
such as VXR37 and VXR49. Hence the X-ray emission of VXR46 and
VXR56 is   
consistent with  arising from a K or M late-type companion.  
Our finding for IC 2391 (at an age of $\sim$ 30 Myr) are compatible 
with the Chandra results for the Pleiades, suggesting that 
late B and A stars are not strong intrinsic X-ray sources \citep{Dan02}.
A recent analysis of two early-type stars in the star-forming region Cha I, 
observed with XMM-Newton \citep{Ste04}, indicates a more complex behavior: 
one star (a Herbig Ae/Be) has X-ray properties indistinguishable from 
lower-mass young stars while the other A-type star is particularly soft, 
a characteristic expected for a wind-driven source. 

Note that, on the basis of ROSAT data, \citet{Pan99} have also found 
that dF  X-ray spectra are softer than those of later-type stars. 
Furthermore  the ACIS observations of NGC 2516 \citep{Mice01} and 
the EPIC observations of the Pleiades \citep{Brig03} and Blanco 1 
\citep{Pi04} show that the X-ray spectra of dF stars tend to be softer 
than those of star of other spectral types.
It is intriguing that this different behaviour occurs for stars over 
a wide range of ages. 
The low coronal temperature of dF stars implies  that these
stars are unable to sustain hot coronae  as the cooler stars do, 
suggesting a different efficiency of the  coronal heating mechanism 
responsible for the X-ray emission.
This could be related to, and is a tracer of, the reduced efficiency of 
the $\alpha - \Omega$ dynamo owing to the convection zone becoming shallower 
at early F spectral type. 

\subsection{Flux and Luminosity determination}
In order to convert the count rates to  flux for cluster members  
detected  with  limited  statistics, we have adopted a MOS constant 
conversion factor of 4.11 $\times 10^{-12}$ erg cm$^{-2}$/count, obtained 
assuming that all cluster members have similar spectra. 
In order to  compute its value, we derived first the pn conversion factor 
from count rates to  flux  as the mean of the ratio $flux/Count~Rate$  
for the stars for which we have  performed a  detailed  spectral analysis. 
The estimated overall uncertainty of the pn conversion factor due to the 
intrinsic spread observed in the sample stars is of the order of $\sim$ 15\%.
Considering that, for our sources, with EPIC/pn we collect about $\sim$ 1.9 
more photons counts than with each MOS, we derived the MOS conversion factor 
appropriate for the source rate  in the summed dataset. 
The X-ray luminosity in the 0.3-3.5 keV band, assuming a cluster distance 
of 162 pc, is given in col. 11 of Table \ref{mem}.

\begin{figure}
\centering
\centerline{\psfig{figure=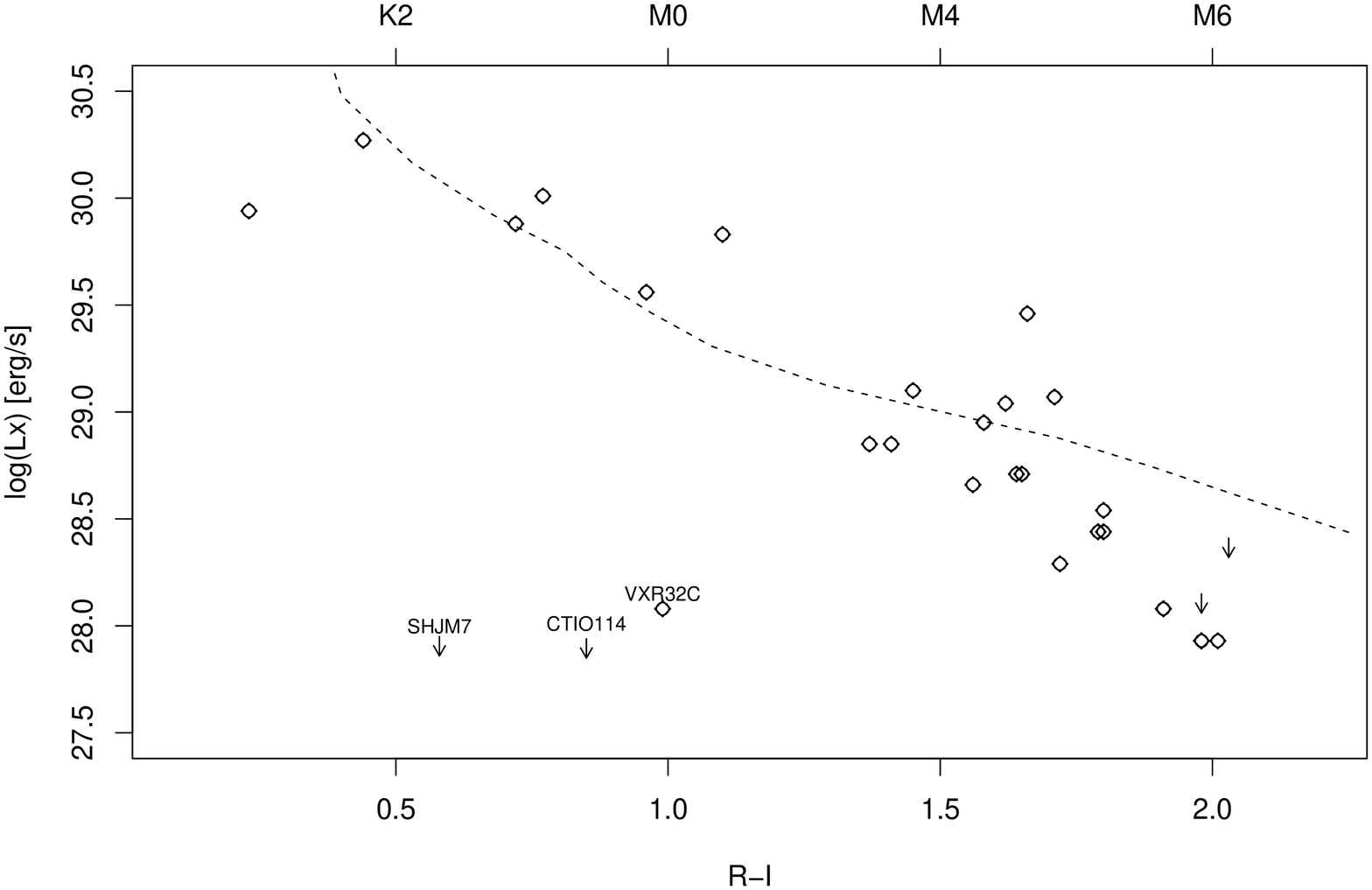,width=9cm,angle=-90}}
\caption{X-ray luminosity versus R-I index color for IC 2391 members. 
Upper limits are marked with downward arrows. The dashed line represents 
the locus of points for which log(L$_x$/L$_{bol}$)=$-3$  for a 30 Myr 
isochrone \citep{Ke94}. } 
\label{LR}
\end{figure}

Fig. \ref{LR} shows X-ray luminosity versus R-I color index for all 
IC 2391 members for which R-I is known. Most of the cluster stars 
emit at the saturation level of L$_x$/L$_{\rm bol} \sim -3$ (dashed 
line in the plot). We note that two stars (VXR37 and VXR47) yield a 
ratio above this level, likely because they have been observed during 
a flare (see Sect. 4), while the two stars SHJM7, CTIO 114 
(for which we have only upper limits for the X-ray emission), 
and the VXR32C+VXR32H+CTIO 96 system are inconsistent with cluster
 membership.  Indeed SHJM7 is reported as a possible cluster member by
 \citet{Stau89} and  as a probable background object by \citet{Pa96}. 
CTIO 114 is also a probable field star. 
For the VXR32C+VXR32H+CTIO 96 system, we equally divided the L$_x$ 
among the three stars since XMM-Newton cannot separate them; however, 
it is possible that VXR32C is not a cluster member and therefore most 
of the emission is due to VXR32H and CTIO 96.

\section{Timing analysis}
We have obtained light-curves in the (0.3-7.8) keV bandpass of the pn 
detected sources. Source and background light-curves were extracted from 
regions described in Sect. 2.4.
In Fig. \ref{ltc} we show source (top) and background (bottom) 
light-curves for bright pn detected cluster  members. 
We clearly observe flare-like variations (e.g. on the dM stars
 VXR37, VXR49), but also short-term \v superimposed on more complex 
variations (VXR38, VXR46, VXR52).
The count rate of VXR49 decreases by a factor $\sim$ 3 in about 20 ksec; 
this behaviour can be interpreted as the decay of a long-lasting flare 
followed by a short-lasting flare. 
Similarly, a first flare followed by a secondary bump triggered by 
the main event has been observed on four stars of the Blanco 1 open 
cluster by \citet{Pi04a}. 
By analysing in detail two of these events on a dG9 star, these authors
interpret the light-curve   in terms of a loop arcade ignited in 
sequence with heating during the decays of both flares.  
See also the detailed modeling of a large flare on Prox Cen \citep{Re04}.

Thanks to the quasi-continuous time coverage, we have unambiguously 
detected  X-ray rotational modulation in VXR45, a fast rotating star 
(P$_{rot}\sim$ 19 ksec) in the supersaturated regime \citep{Mar03}. 
The FOV includes also the fast rotating member VXR47, a M2 star with 
a short photometric period of 22 ksec and {\em v~sin~i} $\sim$ 95 km/s 
\citep{Stau89}. As shown in Fig. \ref{ltc}, the VXR47 light-curve suggests 
the presence of rotational modulation. In order to confirm that the X-ray 
emission may be due to rotation modulation, we have folded the X-ray 
light-curve with the photometric period (see Fig. \ref{ff47}).
Phase-related variability hints  at some rotational modulation but
with irregular variations superposed. 
X-ray rotational modulation could be  present also in the VXR38 and 
VXR44+L33 light curves. In particular the light-curve of VXR38 shows
a smooth modulation on a time scale of $\sim$ 10000 sec that is very 
different from the photometric period of VXR38, namely 2.7 days \citep{Pa96}, 
while for VXR44+L33 there are no photometric periods reported in the literature.  
The detection of rotational modulation implies the presence of non-uniformly 
distributed active regions on these stars. 
Moreover, in VXR45  no strong evidence for spectral variations in the  
soft and hard passbands has been found. 
This is consistent with  the hypothesis that the modulation we observe 
is mainly due to a longitudinal concentration of X-ray-emitting material, 
and that at all times the emission is largely due to the same mixture 
of emitting structures.
A possible scenario is that a large polar region is always present 
while  another region at lower latitude is responsible for the observed
modulation, albeit other scenarios are possible.

We applied the unbinned Kolmogorov-Smirnov (K-S) test to all the 
light-curves of our sample.
Column 10 of Table \ref{mem} reports the results in terms of the
confidence level at which we can reject the hypothesis of a constant source.
A summary of K-S test results is given in Table \ref{KS} and in Fig. \ref{ks.p}:
on short-time scales approximately 46\% of the sources were variable at a 
confidence level $>$99\%, 25\% of them  are not variable (confidence level $<90\%$), 
the remaining 29\% are marginally variable (confidence level between 90\% and 99\%). 
Our results do not depend on count statistics and appear to be an intrinsic property
of the IC 2391 stars.

Combining X-ray data obtained with ROSAT and XMM-Newton 
we can explore long-term variability of the late-type stars of IC 2391. 
Twenty-one stars in our sample have been observed both with ROSAT/PSPC in 1992
and XMM-Newton/EPIC in 2001, and sixteen have been observed both with ROSAT/HRI 
in 1994 and XMM-Newton/EPIC. 
In order to compare our results with  the  ROSAT ones, we have derived 
X-ray luminosities in the (0.2-2.0) keV bandpass.

The comparison of the ROSAT/PSPC luminosities and those of this paper 
indicates that most of the stars have variations less than a factor of 
two  on nine-year time scales, with the exception of VXR40. 
This is a dM star,  variable at a confidence level $>$ 99\% in the PSPC 
observation \citep{Pa96} but not variable in the EPIC observation; 
probably  we are observing an effect due to short-term variability  
during the PSPC observation rather than  truly long-term variability. 
Comparing our data with those collected with ROSAT/HRI we do not find 
evidence of variations on seven-year time scales, only VXR38 has varied 
by a factor larger than 2, but similarly to VXR40 it was variable during 
the EPIC observation and  not in the PSPC one, so probably we are again
in presence of short-term variability.

In summary we have not found evidence of long term variability and such 
variability does not increase by exploring longer time scales. 
Our finding is consistent with a scenario in which stars much younger 
than the Sun, i.e. at ages $<$1 Gyr, do not have long-term cycles or their 
cycle amplitudes  are much smaller than the solar one. Similar indications 
have been found from long-term variability analysis of other young open 
clusters, based on ROSAT, CHANDRA and XMM-Newton observations \citep[e.g.][]
{Gag95, Mar03a, Pi04, Ster95}.
On the other hand, thanks to XMM-Newton, we now have growing evidence that
cycle activity is present in other stars with ages comparable to the solar
one \citep{Fa04}.
This suggest that at very young age the coronal emission does not undergo long
cyclic changes, in agreement with similar results established by the Ca H-K 
monitoring survey \citep[e.g.][]{Ba95}. 

\begin{figure*}[t]
\centerline{\psfig{figure=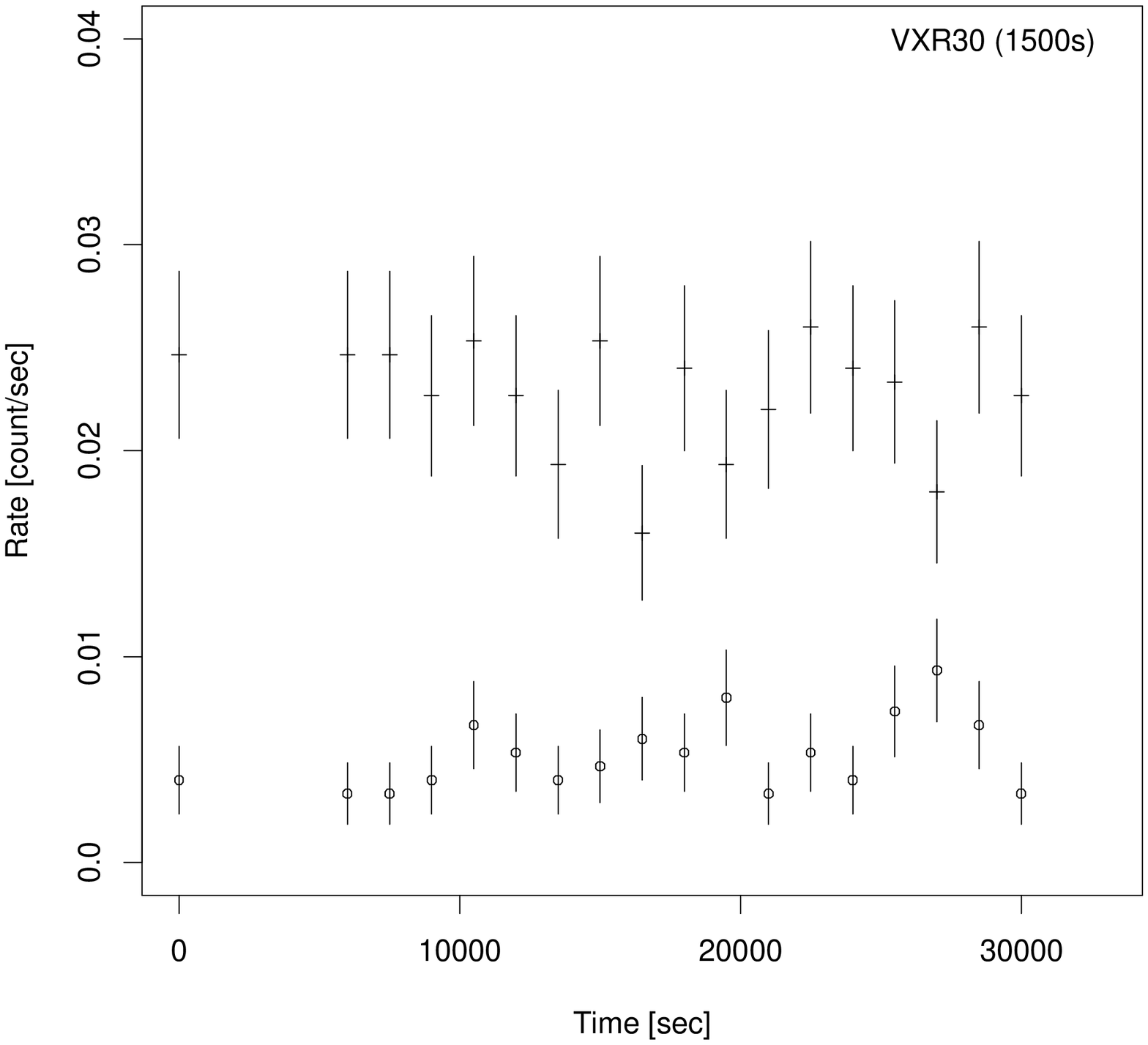,width=7.5cm} \psfig{figure=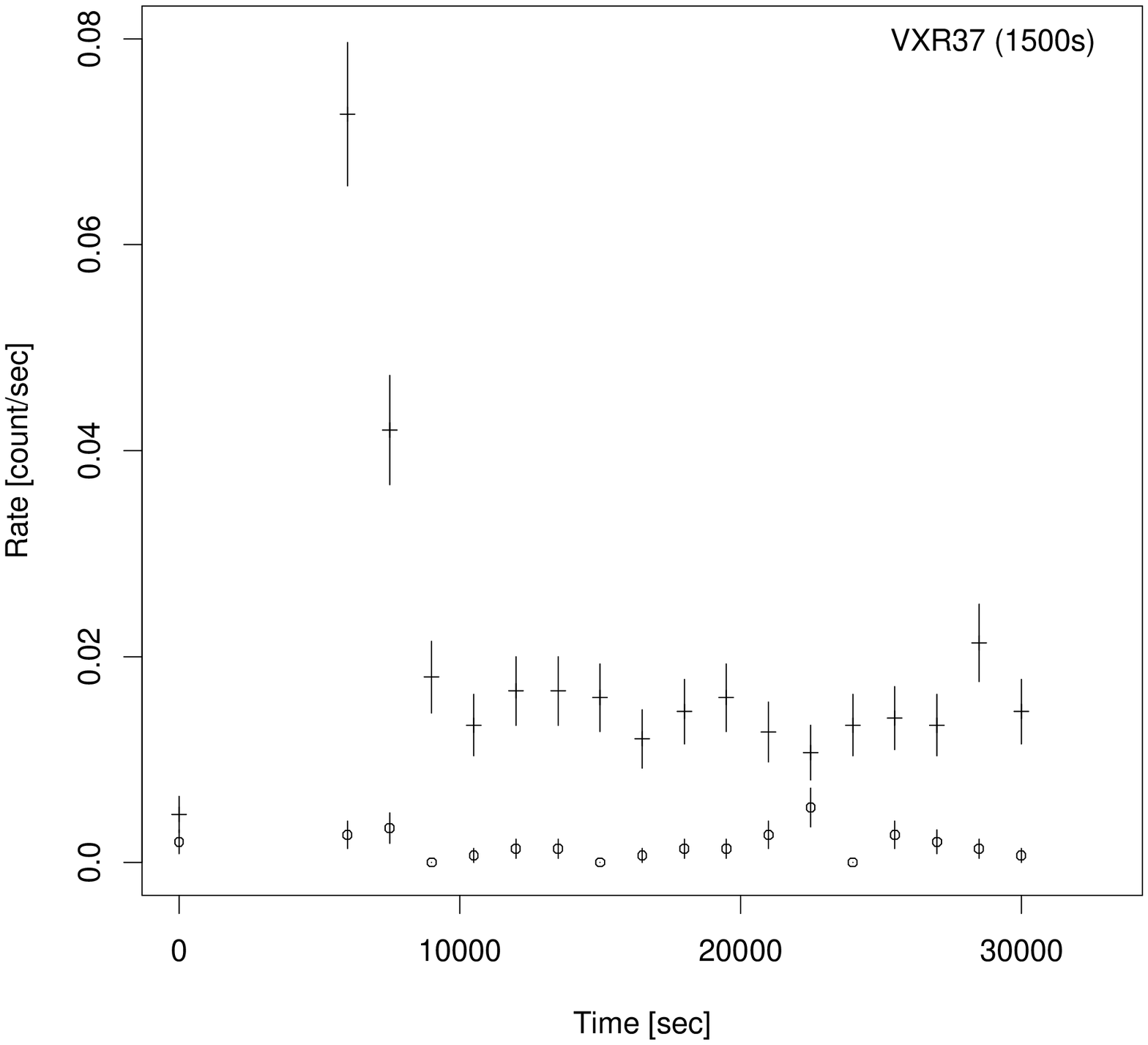,width=7.5cm}}
\centerline{\psfig{figure=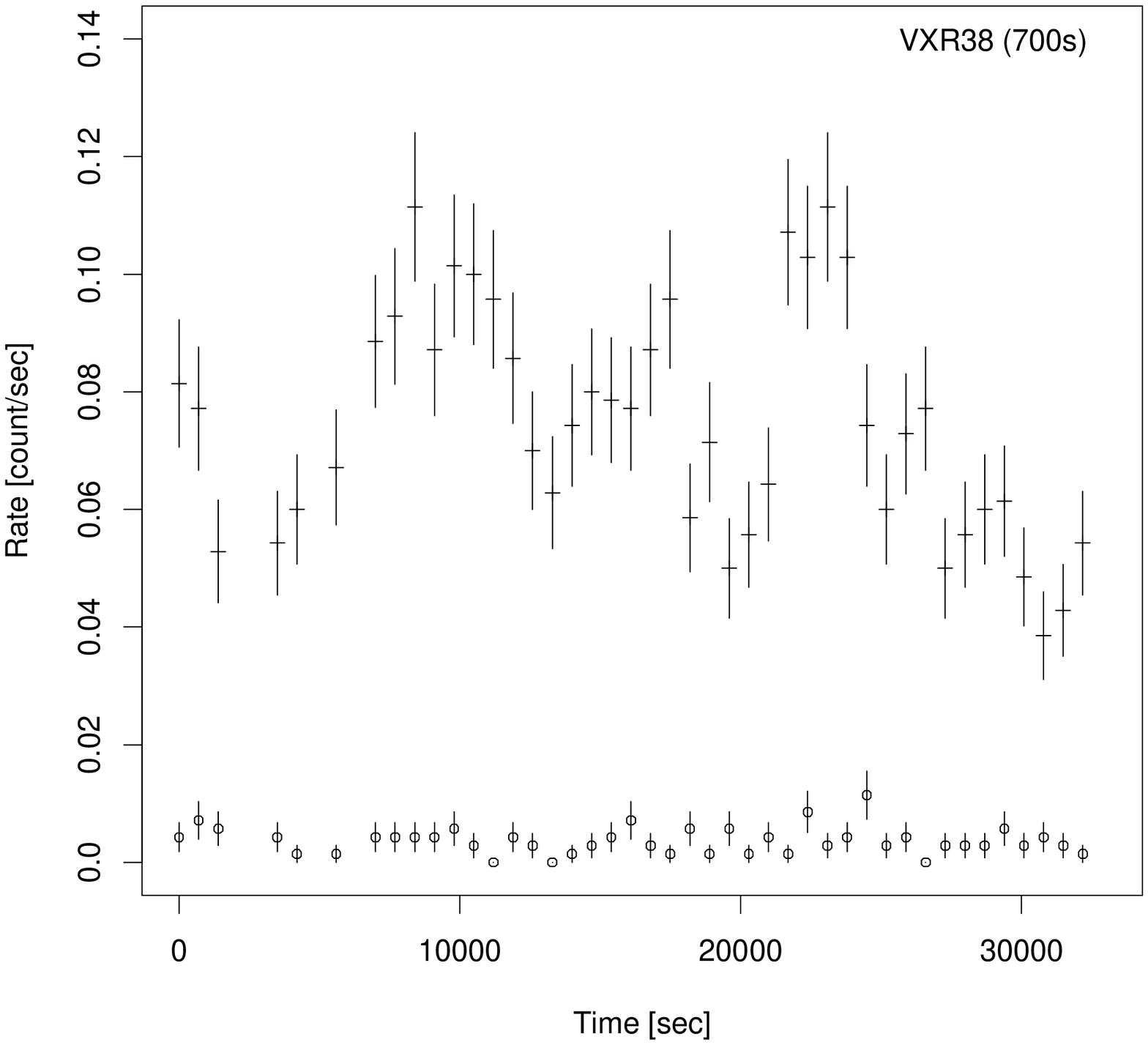,width=7.5cm} \psfig{figure=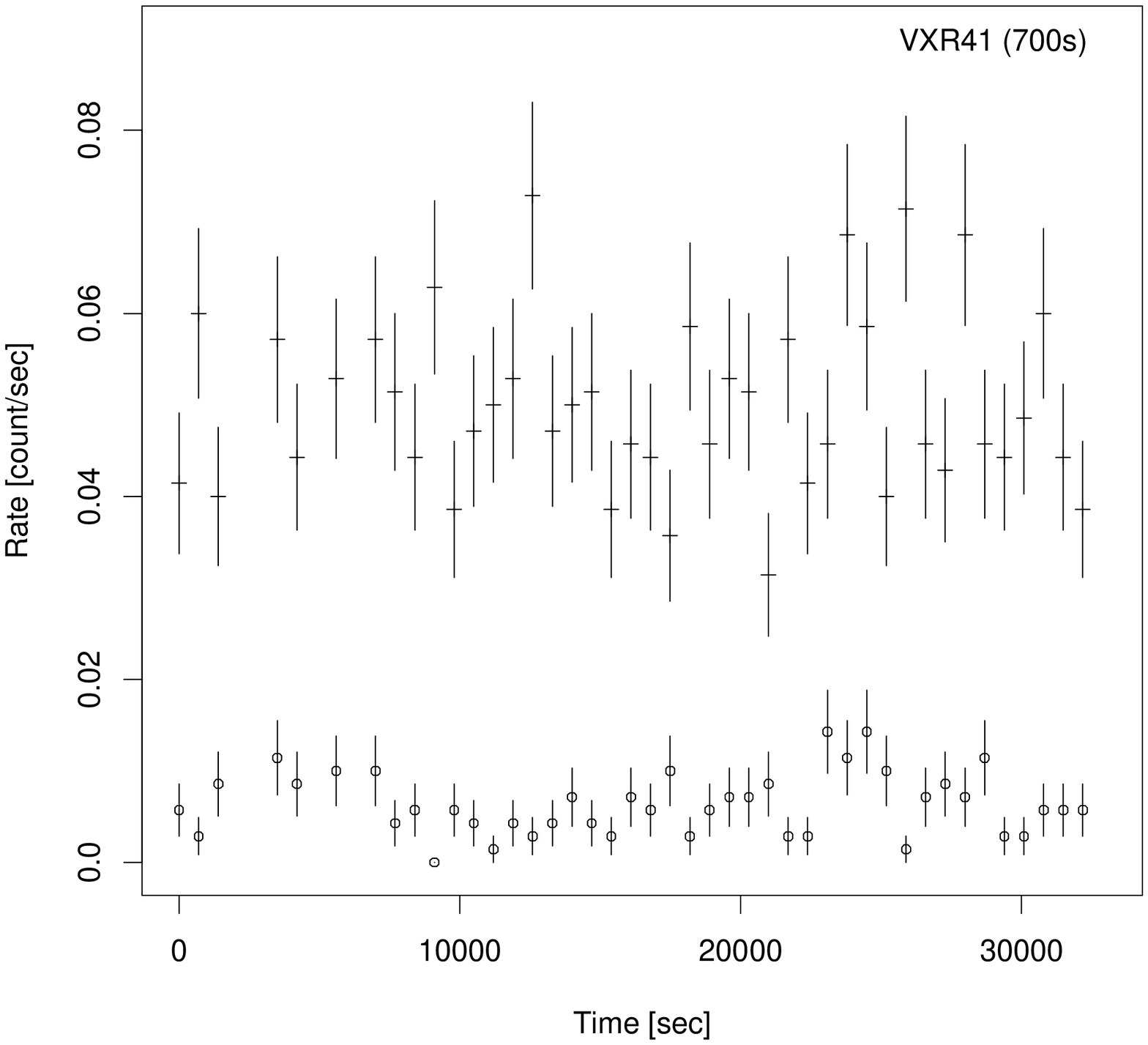,width=7.5cm}}
\centerline{\psfig{figure=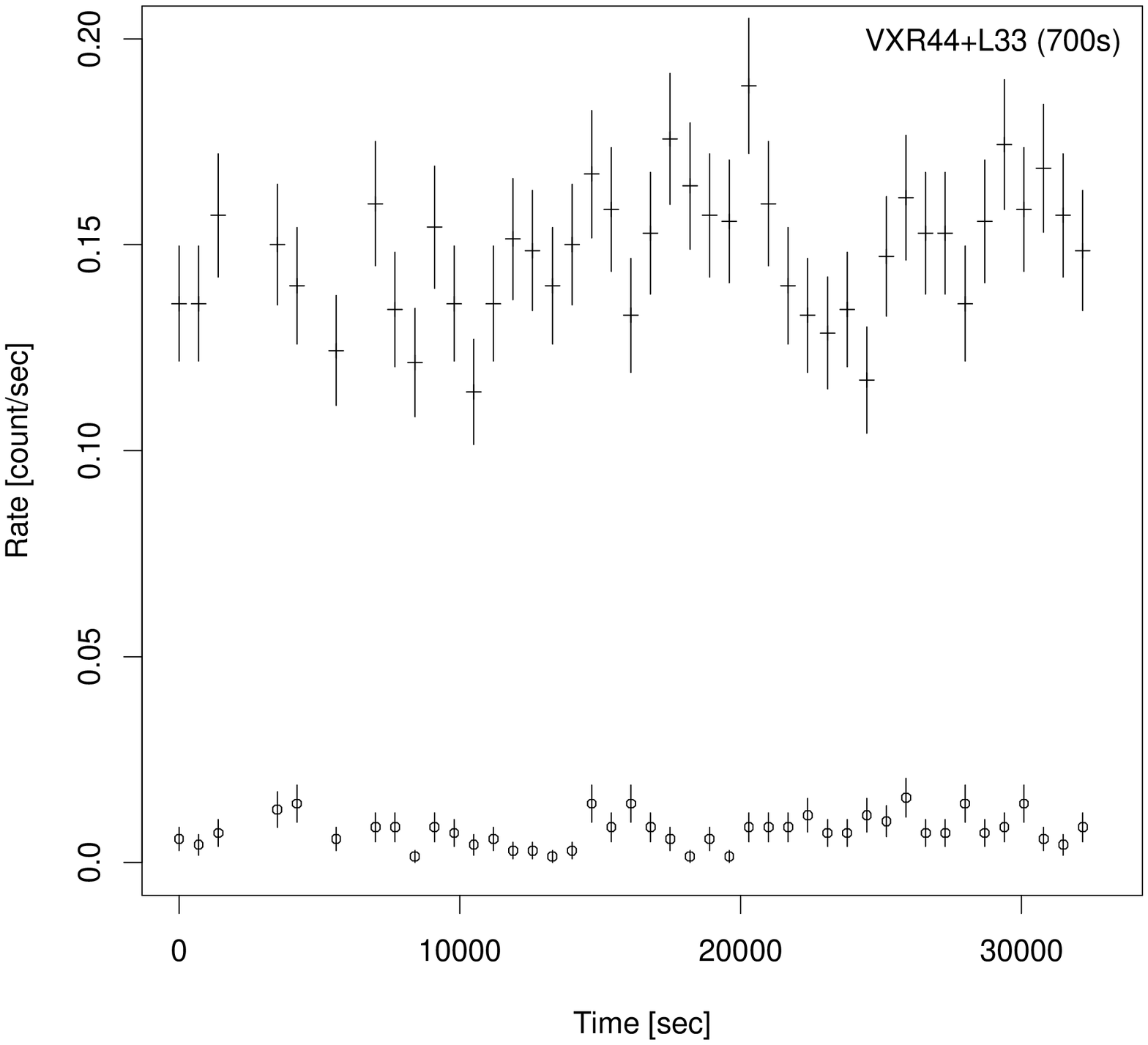,width=7.5cm} \psfig{figure=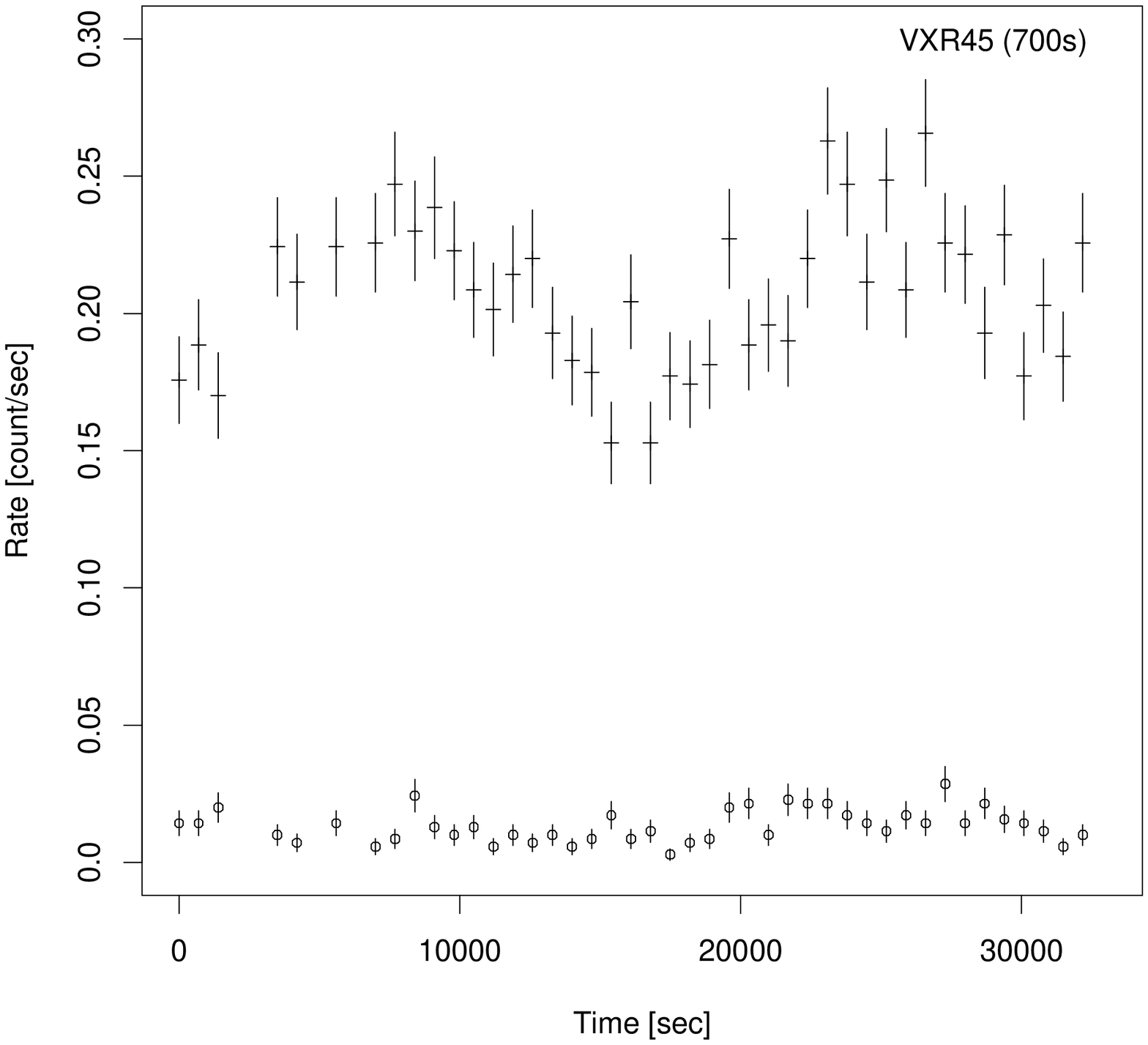,width=7.5cm}}
\caption{Light-curves of X-ray-bright IC 2391 members (top set of crosses in each panel) and background
(bottom set of circles in each panel) in the same XMM-Newton/EPIC FOV.}
\end{figure*}
\addtocounter{figure}{-1}
\begin{figure*}[t]
\centerline{\psfig{figure=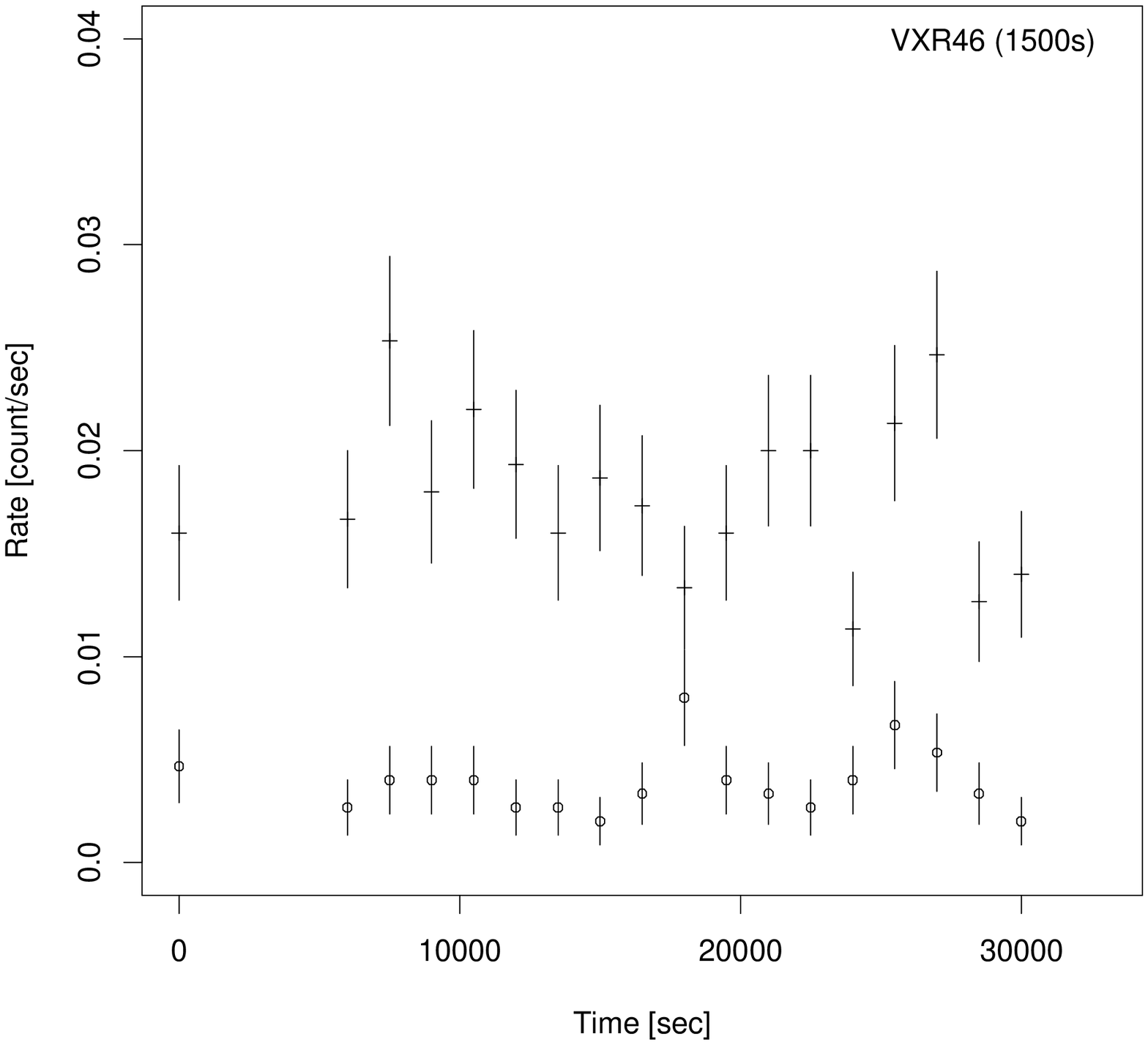,width=7.5cm}  \psfig{figure=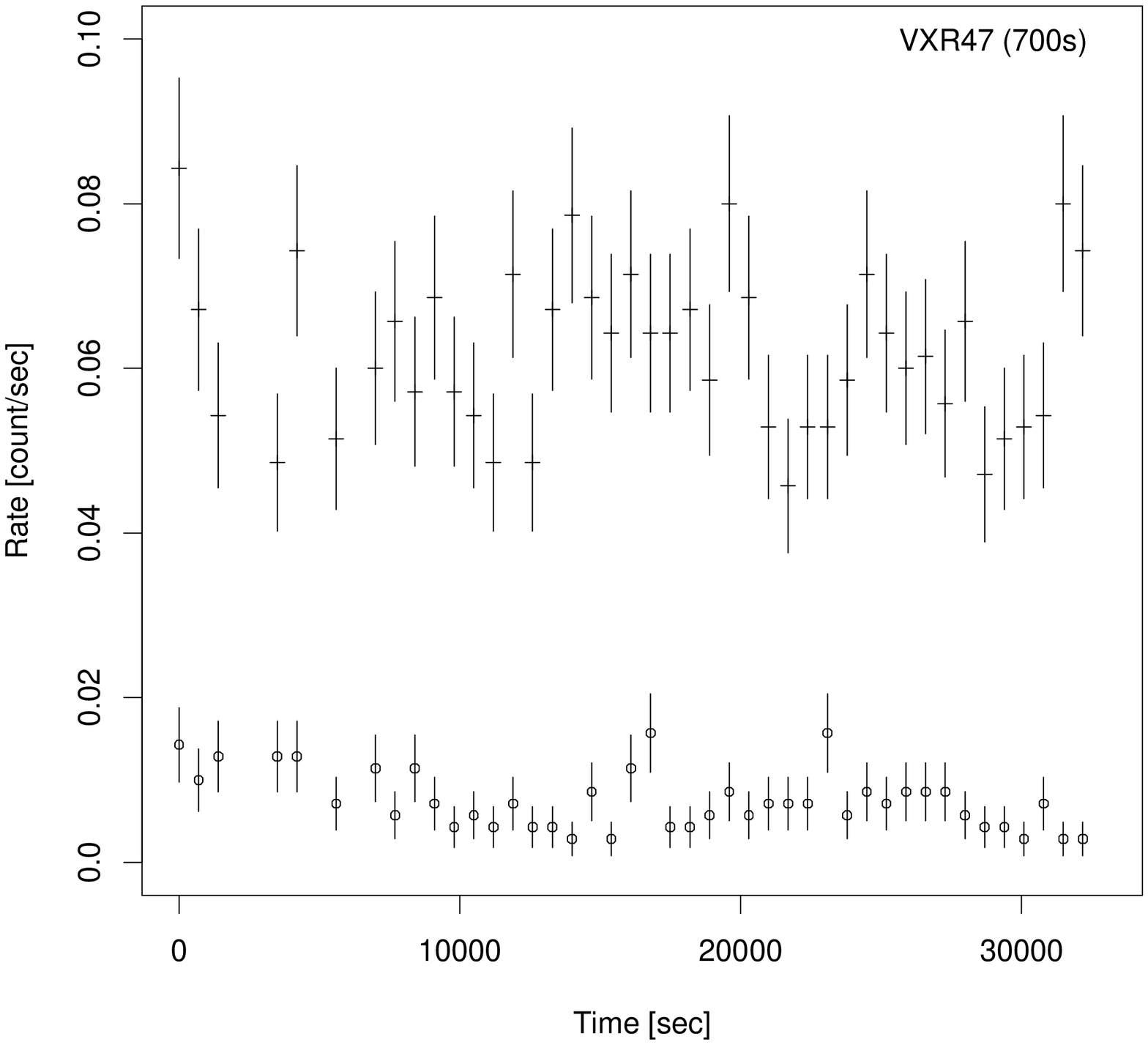,width=7.5cm}}
\centerline{\psfig{figure=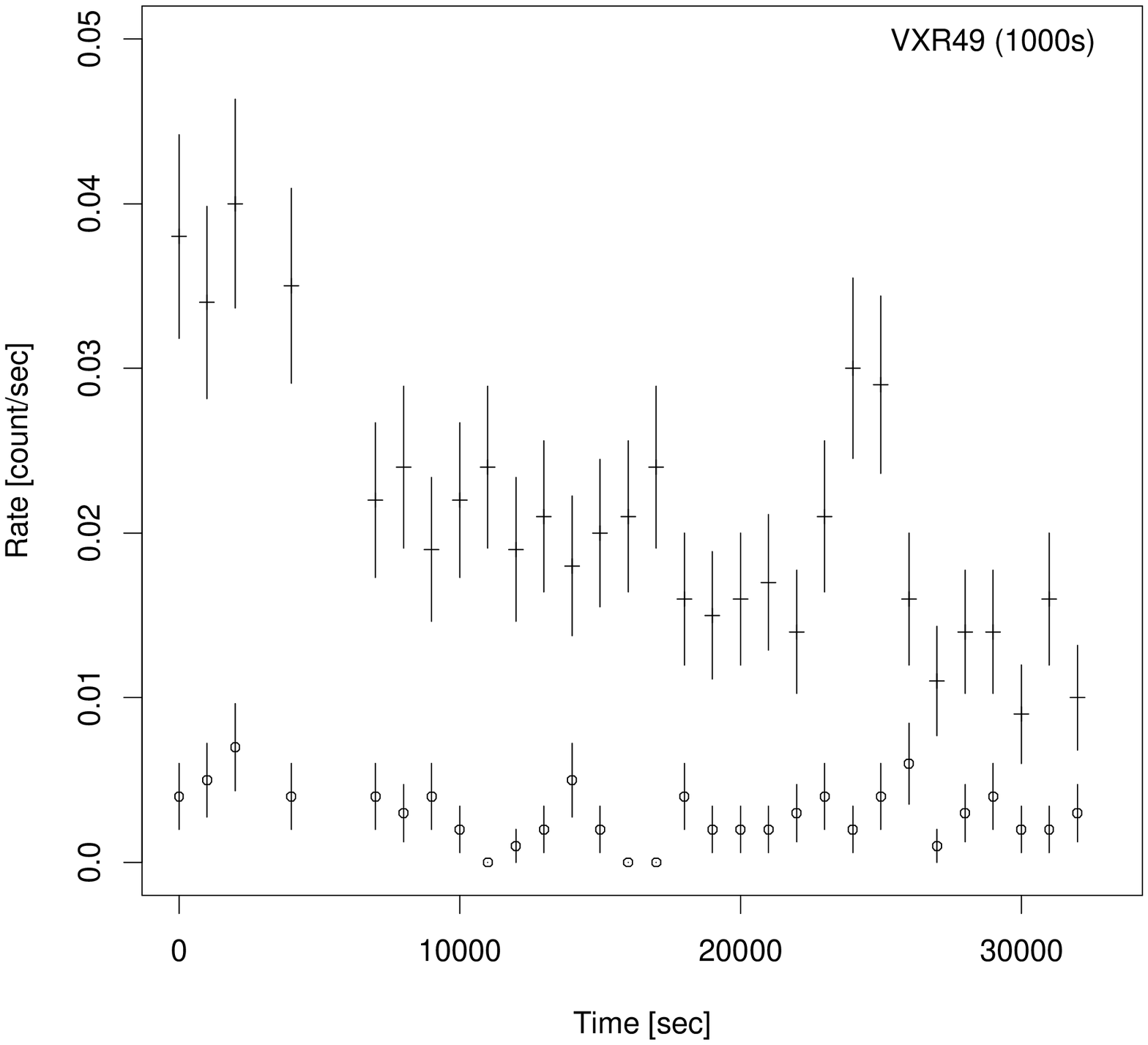,width=7.5cm} \psfig{figure=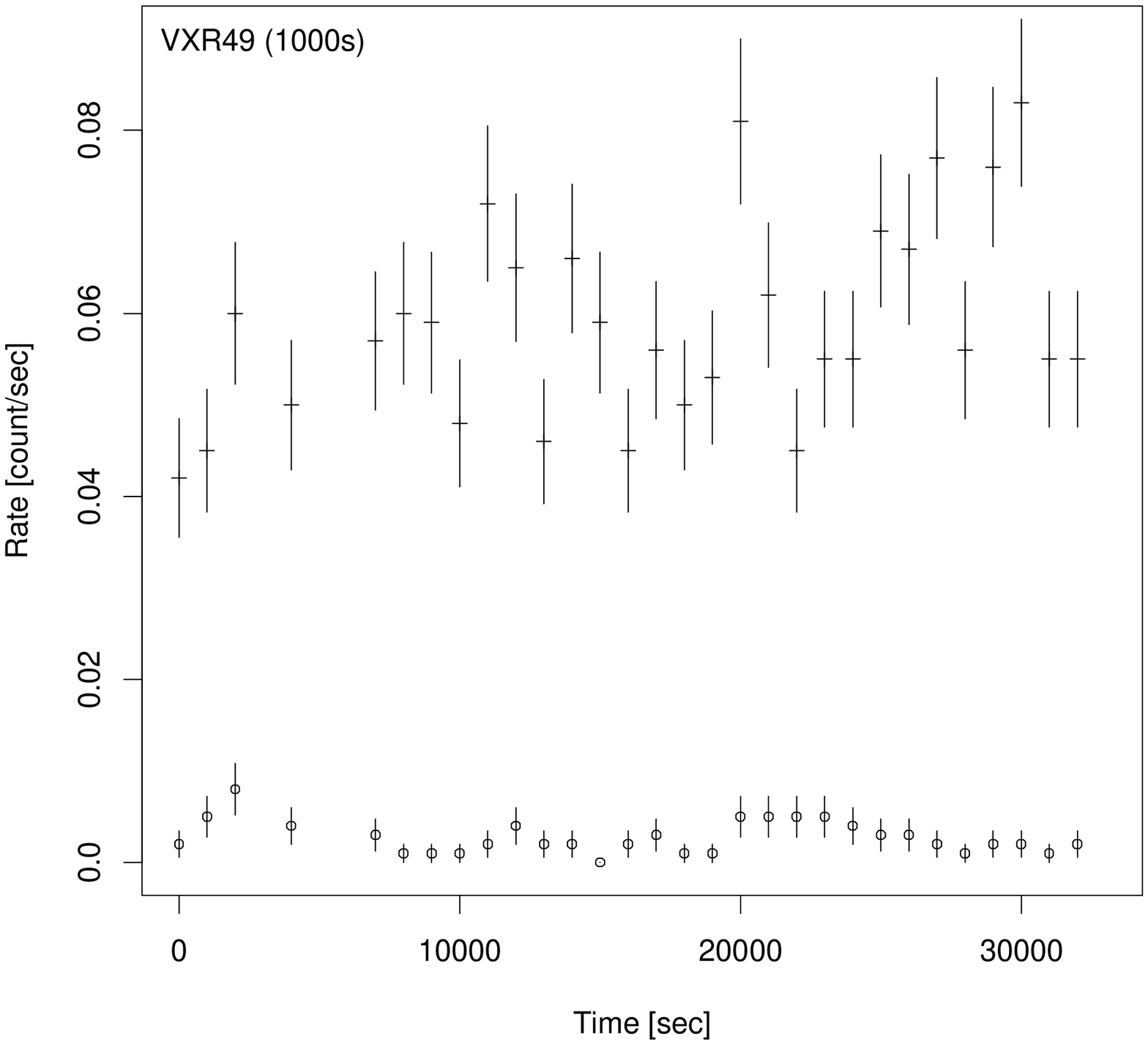,width=7.5cm}}
\centerline{\psfig{figure=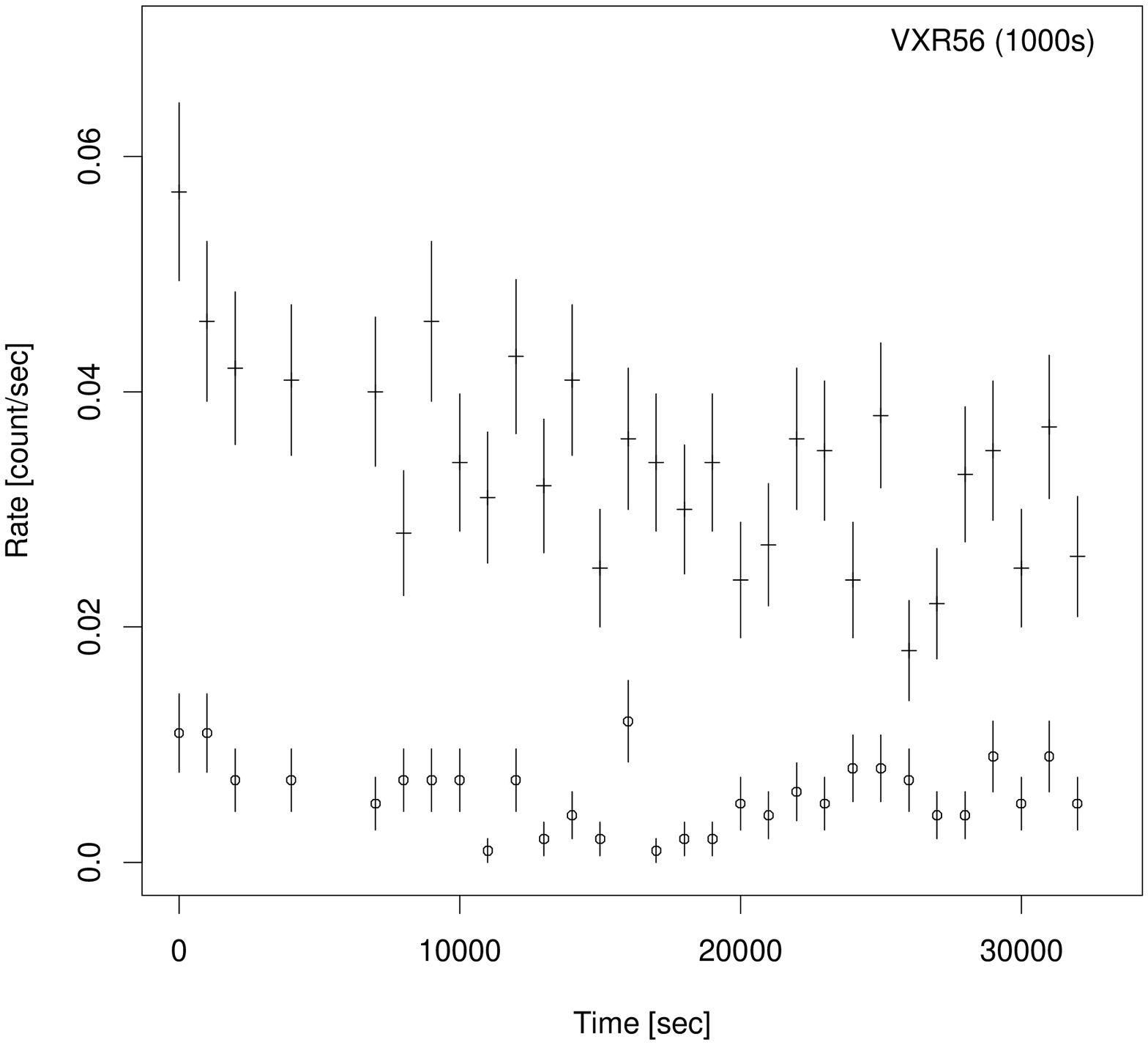,width=7.5cm}}
\caption{Continued}
\label{ltc}
\end{figure*}

\begin{figure}[!h]
  \centerline{\psfig{figure=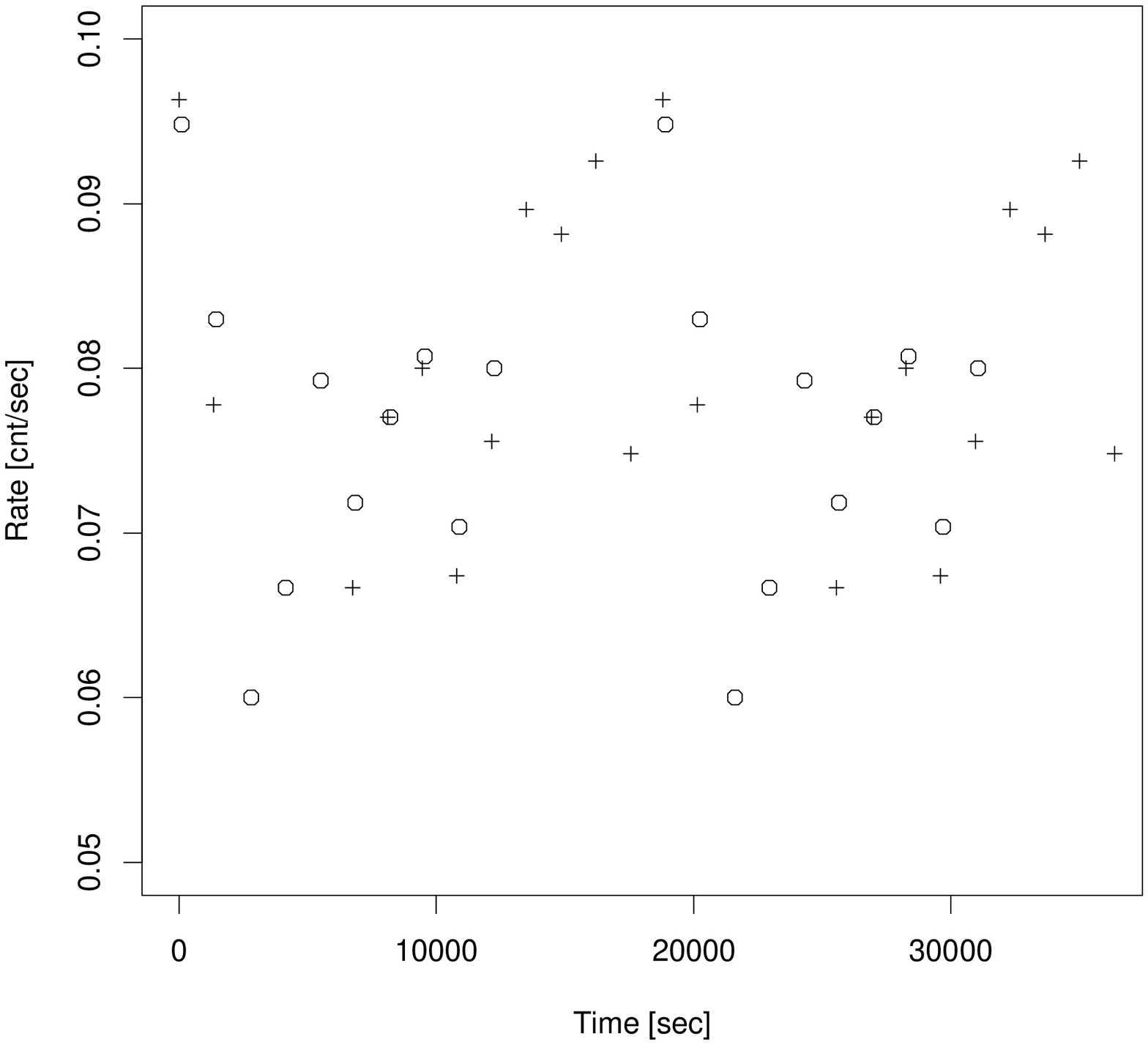,width=9.0cm}}
\caption{Count rate of VXR47 folded with the rotational period vs. phase. 
Circles  indicate measurements for t $<$ 22 ksec (the photometric period) and 
crosses those for t$>$ 22 ksec; all times are referred to the beginning of the
observation. For clarity we show two photometric periods.}
\label{ff47}
\end{figure}

\begin{table}[h]
\caption{Results of the K-S test.}
\bigskip
\centering
\begin{tabular}{|lc|c|c|}
\hline
  &\multicolumn{1}{c|}{Confidence }&\multicolumn{1}{c|}{Number of}\\
&\multicolumn{1}{c|}{level$^1$} & \multicolumn{1}{c|}{identified sources}\\
\hline
  &$>$99\%     & 11 (46\%) \\
  &90\%-99\% & 6 (29\%)   \\
  &$\leq$90\% & 7 (25\%)  \\
\hline
\end{tabular}
\label{KS}

$^1$Confidence level for the rejection of the constant source hypothesis.
\end{table}

\begin{figure}
\centering
\centerline{\psfig{figure=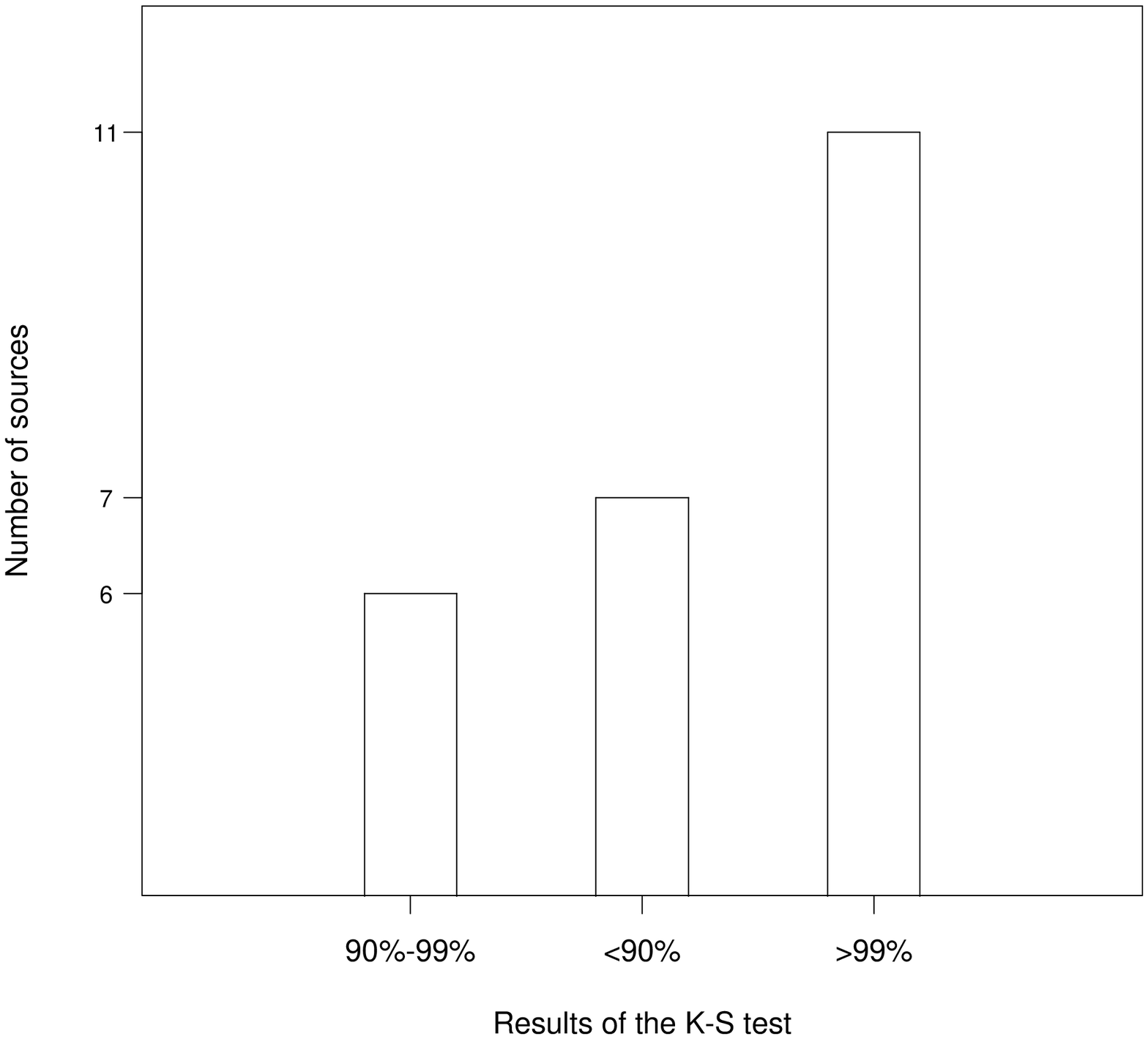,width=7.5cm}}
\caption{Distribution of the confidence level to reject the constant source 
hypothesis for the IC 2391 stars. } 
\label{ks.p}
\end{figure}

\section{Summary}
We have presented the results of the analysis of  an observation of 
the open cluster IC 2391 obtained with XMM-Newton/EPIC camera.

We have detected 31 of the 42 members or probable members of
the cluster in the FOV, corresponding to 24 X-ray sources. 
Furthermore we have computed upper limits at the optical positions 
of undetected cluster members.  
The IC 2391 members having more than 500 pn counts have been the subject
of spectral analysis. 
We find that a 2-T model fits the spectra of G and K type stars while a 
soft 1-T model describes the spectra of F-type stars. 
The spectra of the A1 type star VXR56 and B8.5 type-star VXR46  are 
consistent with the hypothesis that  a later-type companion is responsible
for the  observed X-ray emission.
We have found a rise of the coronal temperature going from F to M stars and 
an indication of subsolar coronal chemical abundances.  

Variability on short time scales is  common among IC 2391 members;
the Kolmogorov-Smirnov test applied to all X-ray photon time series of 
detected cluster members shows that  approximately 46\% of the sources 
are variable at a confidence level greater than 99\%. Furthermore, a very 
fast rotating star of the cluster unambiguously shows  X-ray  rotational 
modulation, while indications of rotational modulation are found in another 
fast rotating star and in two other stars. 
These findings require that in very fast rotating stars the active regions
are typically far from being evenly distributed on the stellar surface.

Comparing our data with published ROSAT data, we find no evidence for 
long-term  cyclic \v with amplitude and cycle period analogous to the solar one. 

\begin{acknowledgements}{This work is based on observations obtained by XMM-Newton, 
an ESA science mission with instruments and contributions directly funded by ESA 
Member States and the USA (NASA). We acknowledge financial support from ASI and MIUR. 
This research has made use of the Open Cluster Database, as provided by C.F. 
Prosser and J.R. Stauffer and which currently may be accessed at 
http://www.noao.edu/noao/staff/cprosser/, or by anonymous ftp to 140.252.1.11, 
cd /pub/prosser/clusters/.  This publication makes use of data products
from the Two Micron All Sky Survey, which is a joint project of the University 
of Massachusetts and the Infrared Processing and Analysis Center/California Institute
of Technology, funded by the National Aeronautics and Space Administration and the
National Science Foundation. We acknowledge the usage of the photographic data 
obtained using The UK Schmidt Telescope and the STScI digitization. 
We wish to thank the referee, Dr. Ramon Garcia Lopez, for his useful comments.}
\end{acknowledgements}
\bibliographystyle{aa}
\bibliography{1525}
\end{document}